\documentclass[a4paper, 11pt]{article}
\pdfoutput=1
\usepackage{jcappub}

\usepackage{amsfonts}
\usepackage{graphicx}
\usepackage{amssymb}
\usepackage{amsmath}
\usepackage{amsthm}
\numberwithin{equation}{section}
\usepackage{cleveref}
\usepackage{bm}
\usepackage[toc, page]{appendix}
\usepackage{breqn}

\title{`Constraint consistency' at all orders in Cosmological perturbation theory}

\author{Debottam Nandi} \author{and S. Shankaranarayanan}
\affiliation{School of Physics, Indian Institute of Science Education
  and Research Thiruvananthapuram (IISER-TVM), India}

\emailAdd{debottam@iisertvm.ac.in}
\emailAdd{shanki@iisertvm.ac.in}

\abstract{We study the equivalence of two --- order-by-order
  Einstein's equation and Reduced action --- approaches to
  cosmological perturbation theory at all orders for different models
  of inflation. We point out a crucial consistency check which we
  refer to as `Constraint consistency' condition that needs to be
  satisfied in order for the two approaches to lead to identical
  single variable equation of motion. The method we propose here is
  quick and efficient to check the consistency for any model including
  modified gravity models. Our analysis points out an important
  feature which is crucial for inflationary model building i.e., all
  `constraint' inconsistent models have higher order Ostrogradsky's
  instabilities but the reverse is not true. In other words, one can
  have models with constraint Lapse function and Shift vector, though
  it may have Ostrogradsky's instabilities. We also obtain single
  variable equation for non-canonical scalar field in the limit of
  power-law inflation for the second-order perturbed variables.}
\notoc
  
\begin{document}

\maketitle

\section{Introduction}
Inflation has now become an integral part of the standard model, that
can eliminate cosmological initial value problems, explain
homogeneities as well as inhomogeneities and observation of
anisotropic Cosmic Microwave Background Radiation
(CMBR)\cite{Komatsu:2008hk}. It is a period of accelerated expansion
in the very early universe and it occurs around $10^{14}$ GeV which is
much remote in time compared to the terrestrial
experiments. Inflationary cosmology has two key theoretical
aspects. One is the approximation schemes employed in solving gravity
equations. The other is the inflationary model building inspired by
particle physics or a fundamental theory of Quantum gravity. The
problem with any theory of gravity is that it is typically highly
non-linear, so one has to rely on approximation schemes to match the
observations. Primarily there exist two formalisms to deal with the
non-linear equations:
\begin{itemize}
\item The separate universe
  approximations\cite{Rigopoulos:2004gr,Rigopoulos:2004ba,Rigopoulos:2005xx}  with either gradient expansion theory or $\Delta N$
  formalism.\cite{Salopek:1990,Sasaki:1998ug,Lyth:2004gb,Lyth:2005fi,Langlois:2005ii,Langlois:2006vv,Sasaki:1995aw}
\item Gauge invariant cosmological perturbation
  theory\cite{Bardeen:1980kt,PhysRevLett.78.1624,Bruni:1996im,Acquaviva:2002ud,Nakamura:2003wk,Maldacena2003,PhysRevD.69.104011,Bartolo:2001cw,Rigopoulos:2002mc,Bernardeau:2002jf,Bernardeau:2002jy,Malik:2003mv, Bartolo:2003bz,Finelli:2003bp,Bartolo:2004if,Enqvist:2004bk,Vernizzi:2004nc,Tomita:2005et,Lyth:2005du, Seery:2005gb, Malik:2005cy, Seery:2006vu,Langlois1994}.
\end{itemize}
The temperature fluctuations as observed in CMB is $\sim 10^{-5}$,
hence it is consistent to use order-by-order perturbation theory to
match with
observations\cite{Bardeen:1980kt,Mukhanov:1990me,Sasaki1986}. In the
first order, one assumes that the perturbed fields are linear. This
implies that the 3-point and higher order correlation functions are
zero. In the second order, the interactions of the first order need to
be included, hence, leading to non-zero 3-point functions. Also it is
widely believed that the detection of these 3-point correlation
functions can reduce the field space of inflationary
models\cite{Bruni:1996im,Maldacena2003,Seery:2005wm}.

With respect to inflationary model building, the proposed theories are
primarily preferred through simplicity. In the case of the canonical
scalar field, the simplest, $60$ e-foldings of inflation require the
potential to be flat, which is in contradiction with particle physics
models\cite{Lyth:1998xn,Lidsey:1995np}. Non-canonical scalar field
model\cite{Armendariz-Picon1999,Garriga1999,Armendariz-Picon2001}
removes the dependence of the potential, however it leads to time
dependence of the speed of perturbations and makes it difficult to be
compared with CMB observations\cite{Phys}. In order to seek more
generalized fields, scalar fields with higher time derivatives in
action are
considered\cite{Kobayashi2010,Kobayashi2011,Nicolis2008,Deffayet2009}. Beside
these, modified gravity models, specifically $f(R)$ lead to
accelerated expansion in the early universe.

There are two mathematical procedures that are currently used in the
literature to study gauge invariant cosmological perturbation theory:
Hamiltonian formulation(ADM formulation)\cite{Langlois1994} and
Lagrangian
formulation\cite{Bardeen:1980kt,PhysRevLett.78.1624,Bruni:1996im,Acquaviva:2002ud,Nakamura:2003wk,Maldacena2003,
  PhysRevD.69.104011,Bartolo:2001cw,Rigopoulos:2002mc,
  Bernardeau:2002jf,Bernardeau:2002jy,Malik:2003mv, Bartolo:2003bz,
  Finelli:2003bp,
  Bartolo:2004if,Enqvist:2004bk,Vernizzi:2004nc,Tomita:2005et,
  Lyth:2005du, Seery:2005gb, Malik:2005cy, Seery:2006vu}. Since
gravity and matter are coupled to each other, one can write the full
action and vary the action with respect to metric and matter fields to
obtain general equations of motion (e.g., Einstein's equation in
General relativity). Those equations can be expanded in terms of
perturbed variables (metric and field variables) and one can write
down equations in the perturbation theory\cite{Lifshitz:1963ps}. In
the action formalism\cite{Lukash:1980iv}, the action is expanded to
the required perturbed order in terms of the perturbed variables
(metric and field variables) and then varied the action with respect
to these variables. For example, to obtain perturbed equations in the
first order, we need to expand the action to second order and vary the
action with respect to the first order perturbed variables. This
formalism can be extended to obtain perturbed equations of motion up
to any order. In the reduced action formalism, constraint variables
are replaced in the action using constraint equations so that we can
rewrite the action only in terms of dynamical variables.

Since the matter fields (Non-canonical \& Galilean scalar field model)
and Gravity are highly non-linear, it is not clear whether the two
approaches, i.e., Einstein's equations writing in order-by-order
perturbation theory and action/reduced action formalism, lead to the
same equations of motion. In Ref.\cite{Appignani2010a}, it was shown
that when the metric perturbations are frozen then the two approaches
do not, in general, lead to the same expressions. In this work we
address the issue by including the metric perturbations in the theory.

In the next section, we study higher order cosmological perturbation
theory for a single scalar field minimally coupled to gravity and show
the equivalence of the two approaches at all orders. We point out a
crucial and novel consistency check which we refer to as `constraint
consistency' condition that needs to be verified. We also show that
this provides a fast and efficient way to check the consistency and
apply it to minimally coupled non-canonical scalar field.

In section \ref{sec3}, we apply the `constraint consistency' condition
to many inflationary models that are proposed in the literature. First
we check the theory with higher derivative Lagrangian models minimally
coupled to gravity. Then we extend the procedure to other different types
of models like modified gravity models and modified gravity with
higher order matter Lagrangian. Appendix \ref{App:Appendix A} contains
some of the derived expressions used in section \ref{sec2} and in
Appendix \ref{App:Appendix B}, we obtain a single variable equation of
motion for non-canonical scalar fields in terms of second order
perturbed variables.

In this work, the number of space-time dimensions is 4 and the metric
signature we use is $[-, +, +, +],~ \kappa = 8 \pi G,~ c = 1$.

\section{Consistency of Higher order perturbations in two different
  approaches}\label{sec2}
The action for gravity sourced by a single, non-minimally coupled
scalar field ($\varphi$) is,
\begin{equation}
\label{act}
\mathcal{S} = \int d^4x\sqrt{-g} \left( \frac{R}{2 \kappa} + 
  \mathcal{L}_m \right)
\end{equation}
where  
\begin{equation}\label{boxx}
  \mathcal{L}_m = P(X, \varphi) + G(X, \varphi) \Box \varphi,~~~ ~X \equiv 
\frac{1}{2} g^{\mu \nu} \partial_{\mu}{\varphi} \partial_{\mu}{\varphi}, ~~~ 
\Box \equiv -\frac{1}{\sqrt{-g}}\partial_{\mu}\{\sqrt{-g} g^{\mu \nu} \partial_{\nu}\}
\end{equation}
is the Lagrangian for the Galilean field which is the most general
scalar field model leading to second order equations of
motion. Varying the action with respect to metric gives Einstein's
equation,
\begin{equation}
\label{eine}
R_{\mu \nu} - \frac{1}{2} g_{\mu \nu}~ R = \kappa~ T_{\mu \nu}, 
\end{equation}
where the stress tensor $T_{\mu \nu}$ is,
\begin{equation}
\label{TG}
\begin{split}
T_{\mu \nu} &= g_{\mu \nu} \{ P + G_X g^{\alpha
  \beta} \partial_{\alpha}{X} \partial_{\beta}{\varphi} + G_\varphi
g^{\alpha
  \beta} \partial_{\alpha}{\varphi} \partial_{\beta}{\varphi}\} \\
 & - \{P_X + 2 G_\varphi + G_X \Box
\varphi\} \partial_{\mu}{\varphi} \partial_{\nu}{\varphi} - 2
G_X \partial_{\mu}{X} \partial_{\nu}{\varphi}
\end{split}
\end{equation}

For simplicity and to obtain the physical features, we consider only
single scalar field theory minimally coupled to the gravity. In
Section \ref{fr}, we look at modified gravity models. Variation of the
action (\ref{act}) with respect to the scalar field `$\varphi$' leads
to the following equation of motion,
\begin{equation}
\label{EG}
\begin{split}
  &\{2 G_\varphi - 2 X G_{X \varphi} + P_X\} \Box \varphi - \{P_{XX} +
  2 G_{X \varphi}\}\partial_{\mu}{\varphi} \partial^{\mu}{\varphi} -
  2X
  \{G_{\varphi} + P_{X \varphi} \} + P_\varphi\\
  & - G_X \{ \varphi_{, \mu \nu} \varphi_{,}^{\mu \nu} - \{\Box
  \varphi\}^2 + R_{\mu
    \nu}\partial^{\mu}{\varphi} \partial^{\nu}{\varphi}\} - G_{XX}
  \{\partial_{\mu}X \partial^{\mu}X +
  \{\partial_{\mu}{\varphi} \partial^{\mu}X\}\Box \varphi\}\} = 0
\end{split}
\end{equation}

As one can see, although the Lagrangian is of the form, $\mathcal{L}_m
= \mathcal{L}_m ( \varphi,\partial{ \varphi}, \partial^2 {\varphi}, t)
$, i.e., it contains higher time derivatives of the scalar field but
equations of motion are second order, thus does not suffer from
Ostrogradsky's instability\citep{Ostro}. With $G(X, \varphi) = 0$ the
field becomes non-canonical. Further fixing $ P = - X - V(\varphi)$,
where $V(\varphi)$ is the potential, the Lagrangian corresponds to
canonical scalar field.

The four-dimensional line element in the ADM form is given by,
\begin{eqnarray}
ds^2 &=& g_{\mu \nu} dx^{\mu} dx^{\nu} \nonumber \\
\label{line}
&=& -(N^2 - N_{i} N^{i} ) d\eta^2 + 2 N_{i} dx^{i} d\eta + \gamma_{i j} dx^{i} dx^{j},
\end{eqnarray}
where $N(x^{\mu})$ and $N_i(x^{\mu})$ are Lapse function and Shift
vector respectively, $\gamma_{i j}$ is the 3-D space metric. Note
that, in the case of Galilean model, $N(x^{\mu})$ and $N_i(x^{\mu})$
are the gauge constraints and variation of action (\ref{act}) with
respect to those lead to Hamiltonian and Momentum constraints,
respectively.

Action (\ref{act}) for the line element (\ref{line}) takes the form,
\begin{equation}
\label{AA}
\mathcal{S} = \int d^4x \sqrt{-g} \Big\{ \frac{1}{2 \kappa}
\left(^{(3)}R + K_{i j} K^{i j} - K^2\right) + \mathcal{L}_m \Big\}
\end{equation}
where $K_{i j}$ is extrinsic curvature tensor and is given by
\begin{eqnarray}
  && K_{i j} \equiv \frac{1}{2N} \Big[ \partial_{0}{\gamma_{i j}} - N_{i|j} - N_{j | i} \Big] \nonumber \\
  && K \equiv \gamma^{i j} K_{i j} \nonumber
\end{eqnarray}

Perturbatively expanding the metric and the scalar field about the
flat FRW spacetime, we get,

\begin{eqnarray}
  && g_{0 0} = - a^2(1 + 2 \epsilon \phi_1 + \epsilon^2 \phi_2 + ...) \\
  && g_{0 i} \equiv N_{i} = a^2 (\epsilon \partial_{i}{B_1} + \frac{1}{2} \epsilon^2 \partial_{i}{B_2} + ...) \\
  && g_{i j} =a^2 \{(1 - 2 \epsilon \psi_1 - \epsilon^2 \psi_2 -...)\delta^{i j} +  2 \epsilon E_{1 |i j} + \epsilon^2 E_{2 |i j} + ...\}\\
  &&\varphi = \varphi_0 + \epsilon \varphi_1 + \frac{1}{2} \epsilon^2 \varphi_2+ ...
\end{eqnarray}
where `$\epsilon$' denotes the order of the perturbation. Note that we
have ignored the vector and tensor part of the metric perturbations.
Although in the first order, the scalar, vector and tensor
perturbations decouple, the three types of perturbations are coupled
in higher order. We assume that the vector and tensor contributions
are small and can be neglected at all orders.
  
To determine the dynamics at every order, we need five scalar
functions ($\phi, B, \psi, E$ and $\varphi$) at each order. Since
there are two gauge choices, one can fix two of the five scalar
functions. In this work, we choose flat-slicing gauge, i.e., $\psi =
0, E = 0$ at all orders,
\begin{eqnarray}
  && g_{0 0} =- a^2(1 + 2 \epsilon \phi_1 + \epsilon^2 \phi_2 + ...) \\
  && g_{0 i} \equiv N_{i} = a^2 (\epsilon \partial_{i}{B_1} + \frac{1}{2} \epsilon^2 \partial_{i}{B_2} + ...) \\
  && g_{i j} =a^2 \delta^{i j}\\
  &&\varphi = \varphi_0 + \epsilon \varphi_1 + \frac{1}{2} \epsilon^2 \varphi_2+ ...
\end{eqnarray}
In the next subsection, we obtain the equation of motion of the second
order perturbed quantity in single variable form for non-canonical
scalar field using order-by-order perturbed Einstein's equation. In
subsection \ref{reducedaction}, we use reduced action approach. To
confirm or infirm the result of Ref.\citep{Appignani2010a}, that the
two approaches lead to different results we focus on non-canonical
scalar field, i.e., setting $G(X, \varphi) = 0$.

\subsection{Order-by-Order Einstein's equation approach}
For the background, $g_{\mu \nu} = diag (- a^2, a^2, a^2, a^2)$,
equations (\ref{eine}) and (\ref{EG}) lead to,

\begin{eqnarray}
\label{Back1}
 &&- \frac{\kappa}{3}  (P_X {\varphi^{\prime}_0}^{2} + P a^2)  = {\mathcal H}^{2} \\
 \label{Back2}
&&- 2 \frac{a^{\prime \prime}}{a} +   \mathcal{H}^2 = \kappa P a^2 \\
\label{Back3}
&& P_X {\varphi_0^{\prime \prime}} - P_{XX} {\varphi_0^{\prime \prime}}
{\varphi_0^{\prime}}^{2} a^{-2} + P_{X\varphi}
{\varphi_0^{\prime}}^{2} + 2 P_X {\varphi_0^{\prime}} \mathcal{H} +
P_{XX} \mathcal{H} {\varphi_0^{\prime}}^{3} a^{-2} + P_{\varphi}
a^{2} = 0
\end{eqnarray}

Equations (\ref{Back1}) and (\ref{Back2}) are zeroth order 0-0 and i-j
Einstein's equations where as equation (\ref{Back3}) is the zeroth
order equation of motion of the scalar field. Similarly, the first
order 0-0, 0-i Einstein's equations and equation of motion of the
perturbed scalar field are,
\begin{eqnarray}
\label{P1}
&&\mathcal{H} \nabla^2{B_1} = \frac{\kappa}{2} ( P_X \phi_1
{\varphi_0^{\prime}}^{2} + 2 P a^2 \phi_1 + 
P_X \varphi_0^{\prime} \varphi_1^{\prime} + P_{XX}
\phi_1 {\varphi_0^{\prime}}^{4} a^{-2} 
- P_{XX} \varphi_1^{\prime} {\varphi_0^{\prime}}^{3} a^{-2}
+ \nonumber \\
&& ~~~~~~~~~~~~~P_{X \varphi} {\varphi_0^{\prime}}^{2} \varphi_1 
+ P_\varphi \varphi_1 a^2 )\\
\label{P2}
&&\mathcal{H} \phi_1= - \frac{\kappa}{2}P_X \varphi_0^{\prime} \varphi_1
\end{eqnarray}

\begin{equation}
\label{P3}
\begin{split}
&- P_X {\varphi_1^{\prime \prime}} a^{2} - P_{XX}
{\phi_1^{\prime}}{\varphi_0^{\prime}}^{3} + P_{XX} {\varphi_1^{\prime
    \prime}} {\varphi_0^{\prime}}^{2} - P_{XX\varphi} \phi_1
{\varphi_0^{\prime}}^{4} 
+ P_{XX\varphi} {\varphi_1^{\prime}}
{\varphi_0^{\prime}}^{3} -P_{\varphi} \phi_1 a^{4} \\
&- P_{\varphi \varphi} a^{4} \varphi_1 + P_X \phi_1 {\varphi_0^{\prime \prime}}
a^{2} + P_X \nabla^2{\varphi_1} a^{2} + P_X {\phi_1^{\prime}}{\varphi_0^{\prime}} a^{2} - 2 P_X {\varphi_1^{\prime}} \mathcal{H}
a^2 - 4 P_{XX} \phi_1 {\varphi_0^{\prime
    \prime}}{\varphi_0^{\prime}}^{2} \\
  &+ 3 P_{XX} {\varphi_0^{\prime}}
{\varphi_1^{\prime}}{\varphi_0^{\prime \prime}} + P_{XX\varphi}
{\varphi_0^{\prime \prime}}{\varphi_0^{\prime}}^{2} \varphi_1 -
P_{X\varphi} {\varphi_0^{\prime}} {\varphi_1^{\prime}} a^{2} -
P_{X\varphi} {\varphi_0^{\prime \prime}} a^{2} \varphi_1 -P_{X\varphi\varphi} {\varphi_0^{\prime}}^{2} a^{2} \varphi_1\\
& + 2 P_X
\phi_1 {\varphi_0^{\prime}} \mathcal{H} a^2 + P_X {\varphi_0^{\prime}}
\nabla^2{B_1} a^{2} + P_{XX} \phi_1 \mathcal{H}
{\varphi_0^{\prime}}^{3} - P_{XX} {\varphi_1^{\prime}}\mathcal{H}
{\varphi_0^{\prime}}^{2} - P_{XXX} \phi_1 \mathcal{H}
{\varphi_0^{\prime}}^{5} a^{-2} \\
&+ P_{XXX} \phi_1 {\varphi_0^{\prime
    \prime}}{\varphi_0^{\prime}}^{4} a^{-2} + P_{XXX}
{\varphi_1^{\prime}}\mathcal{H}{\varphi_0^{\prime}}^{4} a^{-2} -
P_{XXX}{\varphi_1^{\prime}}{\varphi_0^{\prime
    \prime}}{\varphi_0^{\prime}}^{3} a^{-2} - P_{XX\varphi}
\mathcal{H} {\varphi_0^{\prime}}^{3} \varphi_1\\
& - 2 P_{X\varphi}
{\varphi_0^{\prime}}\mathcal{H} \varphi_1 a^2 = 0
\end{split}
\end{equation}

Note that, there are no $\phi_1^{\prime \prime}$ and $B_1^{\prime
  \prime}$ terms in the above three equations and equations (\ref{P1})
and (\ref{P2}) are, as expected, the constraint equations
corresponding to Lapse function and Shift vector. Hence, $\phi_1$ and
$B_1$ are constraints and we can eliminate them from the first order
equation of motion of the scalar field (\ref{P3}). In first order,
single variable equation for non-canonical scalar field in terms of
Mukhanov-Sasaki variable ($v$) is,
\begin{equation}
\label{S1}
v^{\prime \prime} - c_s^2 \nabla^2 v - \frac{z^{\prime \prime}}{z} v = 0
\end{equation}
where,
\begin{equation}
\label{vzc}
v \equiv a \varphi_1,~~~ z \equiv \frac{a \varphi_0^\prime}{\mathcal{H}},~~~ 
c_s^2 \equiv \frac{P_X}{P_X + 2X P_{XX}}
\end{equation}

Similarly, perturbed second order 0-0 and 0-i Einstein's equations for
non-canonical scalar fields at second order are, {\small{
\begin{equation}
\label{G002}
\begin{split}
&- 4 \phi_1 \mathcal{H} \nabla^2{B_1} + \mathcal{H} \nabla^2{B_2} - 2
\delta^{i j} \mathcal{H} \partial_{i}{B_1} \partial_{j}{\phi_1} -
\frac{1}{2} \nabla^2{B_1} \nabla^2{B_1} + \frac{1}{2} {\delta}^{i j}
{\delta}^{k l} {\partial}_{i k}{B_1} {\partial}_{j l}{B_1} \\
& + \kappa (
- \frac{1}{2} P_X {\delta}^{i j} {\partial}_{i}{B_1}
{\partial}_{j}{B_1} {\varphi_0^{\prime}}^{2} - P {\delta}^{i j}
{\partial}_{i}{B_1} {\partial}_{j}{B_1} a^2 - \frac{1}{2} P_X \phi_2
{\varphi_0^{\prime}}^{2} - P \phi_2 a^2 + 2 P_X \phi_1^{2}
{\varphi_0^{\prime}}^{2} + 4 P \phi_1^{2} a^2 \\
&+ 2 P_X \phi_1
\varphi_0^{\prime} \varphi_1^{\prime} + \frac{7}{2} P_{XX} \phi_1^{2}
{\varphi_0^{\prime}}^{4} a^{-2} - 5 P_{XX} \phi_1 \varphi_1^{\prime}
{\varphi_0^{\prime}}^{3} a^{-2} + P_{X \varphi} \phi_1
{\varphi_0^{\prime}}^{2} \varphi_1 - \frac{1}{2} P_X
\varphi_0^{\prime} \varphi_2^{\prime}\\
& - \frac{1}{2} P_X
{\varphi_1^{\prime}}^{2} - \frac{1}{2} P_{XX} \phi_2
{\varphi_0^{\prime}}^{4} a^{-2} - \frac{1}{2} P_{XX} {\delta}^{i j}
{\partial}_{i}{B_1} {\partial}_{j}{B_1} {\varphi_0^{\prime}}^{4}
a^{-2} + 2 P_{XX} {\varphi_0^{\prime}}^{2} {\varphi_1^{\prime}}^{2}
a^{-2} + \frac{1}{2} P_{XX} \varphi_2^{\prime}
{\varphi_0^{\prime}}^{3} a^{-2} \\
& - P_{XX} {\delta}^{i j}
{\partial}_{i}{B_1} {\partial}_{j}{\varphi_1} {\varphi_0^{\prime}}^{3}
a^{-2} - \frac{1}{2} P_{XX} {\delta}^{i j} {\partial}_{i}{\varphi_1}
{\partial}_{j}{\varphi_1}{\varphi_0^{\prime}}^{2} a^{-2} - P_{X
  \varphi} {\varphi_0^{\prime}} {\varphi_1^{\prime}} \varphi_1 -
\frac{1}{2} P_{X \varphi} {\varphi_0^{\prime}}^{2} \varphi_2\\
& -
\frac{1}{2} P_{XXX} \phi_1^{2} {\varphi_0^{\prime}}^{6} a^{-4} 
 + P_{XXX} \phi_1 {\varphi_1^{\prime}} {\varphi_0^{\prime}}^{5} a^{-4} -
\frac{1}{2} P_{XXX}{\varphi_0^{\prime}}^{4} {\varphi_1^{\prime}}^{2}
a^{-4} - \frac{1}{2} P_{X\varphi\varphi} {\varphi_0^{\prime}}^{2}
\varphi_1^{2} - P_{XX\varphi} \phi_1 {\varphi_0^{\prime}}^{4} a^{-2}
\varphi_1 \\
&+ P_{XX\varphi} {\varphi_1^{\prime}}
{\varphi_0^{\prime}}^{3} a^{-2} \varphi_1 
- \frac{1}{2} P_X
{\delta}^{i j} {\partial}_{i}{\varphi_1} {\partial}_{j}{\varphi_1} -
\frac{1}{2} P_\varphi \varphi_2 a^2 - \frac{1}{2} P_{\varphi \varphi}
\varphi_1^{2} a^2) = 0
\end{split}
\end{equation}}}
{\small{
\begin{equation}
\label{G0i2}
\begin{split}
 &- 2 \delta^{j k} \mathcal{H} \partial_{j}{B_1} \partial_{i k}{B_1} +
 \partial_{i}{\Phi_1} \nabla^2{B_1} + 4 \phi_1 \mathcal{H}
 \partial_{i}{\phi_1} - \mathcal{H} \partial_{i}{\phi_2} - \delta^{j
   k} \partial_{j}{\phi_1} \partial_{i k}{B_1}\\
   & +\kappa ( - \frac{1}{2}
 P_X {\varphi_0^{\prime}} \partial_{i}{\varphi_2} - P_X {\varphi_1^{\prime}}
 \partial_{i}{\varphi_1} - P_{XX} \phi_1
 \partial_{i}{\varphi_1}{\varphi_0^{\prime}}^{3} a^{-2} +
 P_{XX}{\varphi_1^{\prime}}
 \partial_{i}{\varphi_1} {\varphi_0^{\prime}}^{2} a^{-2} - P_{X\varphi}
         {\varphi_0^{\prime}} \partial_{i}{\varphi_1} \varphi_1)=0
\end{split}
\end{equation}}}
 and the equation of motion of the scalar field is,
 
 \begin{equation}
\label{eos2}
\begin{split}
&C_X P_X + C_{XX} P_{XX} + C_{XXX} P_{XXX} + C_{XXXX} P_{XXXX} +
C_{XXX\varphi} P_{XXX\varphi}\\
& + C_{XX\varphi} P_{XX\varphi} +
C_{XX\varphi\varphi} P_{XX\varphi\varphi} + C_{X\varphi} P_{X\varphi}
+ C_{X\varphi\varphi} P_{X\varphi\varphi} + C_{X\varphi\varphi\varphi}
P_{X\varphi\varphi\varphi}\\
& + C_{\varphi} P_{\varphi} +
C_{\varphi\varphi} P_{\varphi\varphi} + C_{\varphi\varphi\varphi}
P_{\varphi\varphi\varphi} = 0
\end{split}
\end{equation}
where $C_X, C_{XX}, ...$ are all second order perturbed quantities and
$P_X, P_{XX}, P_{X \varphi}, ...$ are background quantities. The
explicit form of C's are given in Appendix \ref{App:Appendix A}.

It is important to note that the second order equations also do not
contain $\phi_2^{\prime \prime}$ and/or $B_2^{\prime \prime}$. Hence
one can obtain a single variable equation of motion of non-canonical
scalar field at second order. Malik et al obtained the single variable
equation of motion in second order for canonical scalar
field\cite{Malik2007}. Also note that `$\varphi$' evaluated in the
flat-slicing gauge is a gauge invariant quantity and directly related
to comoving curvature perturbation $\mathcal{R}$/curvature
perturbation on uniform-density hypersurfaces $\zeta$
\cite{Malik2009}.

\subsection{Reduced Action approach}\label{reducedaction}
In the reduced action approach, which is now a popular way to
calculate non-gaussianity, one perturbs the field variables ($g_{\mu
  \nu}, \varphi$) in the action and expands the action to the required
order. In other words, one assumes a priori the form of the metric and
the matter variables in the lowest order and expands
order-by-order. For instance, in the case of FRW background, the
action (\ref{AA}) becomes,

\begin{equation}
  ^{(0)}\mathcal{S}^{NC} = \int d^4x \left( P a^{4} - 3 \frac{1}{\kappa} 
    {a^\prime}^{2} \right)
\end{equation}

Varying the above action with respect to metric variable $a(\eta)$ and
$\varphi_0(\eta)$ leads to the equations (\ref{Back2}) and
(\ref{Back3}). Note that, as expected, these two equations are
independent of each other since $a(\eta)$ and $\varphi_0(\eta)$ are
dynamical variables. To obtain first order (in $\epsilon$) equations,
one expands action (\ref{AA}) upto second order of
$\epsilon$. \textit{In general, varying the $n^{th}$ order action with
  respect to $m^{th}$ order perturbed variables leads to $(n -m)^{th}$
  order perturbed equations. It may be worth noting that a given order
  equations of motion can be obtained in several ways, e.g., varying
  first order action with respect to first order variables leads to
  zeroth order equations of motion}.

Expanding the action (\ref{AA}) to the second order, only in terms of
first order variables ($\varphi_1, \phi_1, B_1$), we get {\small
\begin{equation}
\begin{split}
  ^{(2)}\mathcal{S}^{NC} = & \int d^4x\Big(\frac{1}{2} P_X {\delta}^{i
    j} {\partial}_{i}{B_1} {\partial}_{j}{B_1}
  {\varphi_0^{\prime}}^{2} a^{2} - P_X \phi_1^{2}
  {\varphi_0^{\prime}}^{2} a^{2} + P_X \phi_1 {\varphi_0^{\prime}}
  {\varphi_1^\prime} a^{2} - \frac{1}{2} P_X {\varphi_1^\prime}^2
  a^{2} + \\
  &P_X {\delta}^{i j} {\varphi_0^{\prime}} {\partial}_{i}{B_1}
  {\partial}_{j}{\varphi_1} a^{2} + \frac{1}{2} P_X {\delta}^{i j}
  {\partial}_{i}{\varphi_1} {\partial}_{j}{\varphi_1} a^{2} +
  \frac{1}{2} P_{XX} \phi_1^{2} {\varphi_0^{\prime}}^{4} - P_{XX}
  \phi_1 {\varphi_1^\prime} {\varphi_0^{\prime}}^{3} + \frac{1}{2}
  P_{XX} {\varphi_0^{\prime}}^{2} {\varphi_1^\prime}^{2} + \\
  &\frac{1}{2}
  P_{\varphi\varphi} \varphi_1^{2} a^{4} + P_{X\varphi} \phi_1
  {\varphi_0^{\prime}}^{2} a^{2} \varphi_1 - P_{X\varphi}
  {\varphi_0^{\prime}} {\varphi_1^\prime} a^{2} \varphi_1 + P_\varphi
  \phi_1 a^{4} \varphi_1 + \frac{1}{2} P {\delta}^{i j}
  {\partial}_{i}{B_1} {\partial}_{j}{B_1} a^{4} - \frac{1}{2} P
  \phi_1^{2} a^{4} -\\
  & 2 \phi_1 {\delta}^{i j} \frac{1}{\kappa}
  {a^{\prime}} {\partial}_{i j}{B_1} a + \frac{3}{2} {\delta}^{i j}
  \frac{1}{\kappa} {\partial}_{i}{B_1} {\partial}_{j}{B_1}
  {a^{\prime}}^{2} - \frac{9}{2} \frac{1}{\kappa} \phi_1^{2}
  {a^{\prime}} ^{2} + \frac{1}{2} {\delta}^{i j} {\delta}^{k l}
  \frac{1}{\kappa} {\partial}_{i k}{B_1} {\partial}_{j l}{B_1} a^{2} -\\
  &
  \frac{1}{2} {\delta}^{i j} {\delta}^{k l} \frac{1}{\kappa}
  {\partial}_{i j}{B_1} {\partial}_{k l}{B_1} a^{2} \Big)
  \end{split}
\end{equation}}

After integrating by-parts, and dropping off boundary terms, we get,
{\small
\begin{equation}
\label{AP1}
\begin{split}
  ^{(2)}\mathcal{S}^{NC} =& \int d^4x\Big(\frac{1}{2} P_X {\delta}^{i
    j} {\partial}_{i}{B_1} {\partial}_{j}{B_1}
  {\varphi_0^{\prime}}^{2} a^{2} - P_X \phi_1^{2}
  {\varphi_0^{\prime}}^{2} a^{2} + P_X \phi_1 {\varphi_0^{\prime}}
  {\varphi_1^\prime} a^{2} - \frac{1}{2} P_X
  {\varphi_1^\prime}^2 a^{2} + \\
  &P_X {\delta}^{i j} {\varphi_0^{\prime}} {\partial}_{i}{B_1}
  {\partial}_{j}{\varphi_1} a^{2} + \frac{1}{2} P_X {\delta}^{i j}
  {\partial}_{i}{\varphi_1} {\partial}_{j}{\varphi_1} a^{2} +
  \frac{1}{2} P_{XX} \phi_1^{2} {\varphi_0^{\prime}}^{4} - P_{XX}
  \phi_1 {\varphi_1^\prime} {\varphi_0^{\prime}}^{3} +
  \frac{1}{2} P_{XX} {\varphi_0^{\prime}}^{2} {\varphi_1^\prime}^{2} +\\
  & \frac{1}{2} P_{\varphi\varphi} \varphi_1^{2} a^{4} + P_{X\varphi}
  \phi_1 {\varphi_0^{\prime}}^{2} a^{2} \varphi_1 - P_{X\varphi}
  {\varphi_0^{\prime}} {\varphi_1^\prime} a^{2} \varphi_1 + P_\varphi
  \phi_1 a^{4} \varphi_1 + \frac{1}{2} P {\delta}^{i j}
  {\partial}_{i}{B_1} {\partial}_{j}{B_1} a^{4} - \\
  &\frac{1}{2} P \phi_1^{2} a^{4} - 2 \phi_1 {\delta}^{i j}
  \frac{1}{\kappa} {a^{\prime}} {\partial}_{i j}{B_1} a + \frac{3}{2}
  {\delta}^{i j} \frac{1}{\kappa} {\partial}_{i}{B_1}
  {\partial}_{j}{B_1} {a^{\prime}}^{2} - \frac{9}{2} \frac{1}{\kappa}
  \phi_1^{2} {a^{\prime}} ^{2} \Big)
\end{split}
\end{equation}}

Varying action (\ref{AP1}) with respect to $\varphi_1$, we obtain
first order equation of motion of the scalar field same as
(\ref{P3}). Similarly, varying action with respect to $\phi_1$ and
$B_1$ gives same equations as (\ref{P1}) and (\ref{P2}) respectively,
i.e.,
\begin{eqnarray}
  \Big\{\frac{\delta S_2}{\delta \varphi_1}\Big\}_{\phi_1, B_1} &\equiv& 
1^{st}~order~Equation~of~ motion~of~the~scalar~field \nonumber \\
  \Big\{\frac{\delta S_2}{\delta \phi_1}\Big\}_{\varphi_1, B_1} &\equiv& 
1^{st}~order~Hamiltonian~constraint \nonumber \\
  \Big\{\frac{\delta S_2}{\delta B_1}\Big\}_{\varphi_1, \phi_1} &\equiv& 
1^{st}~order~Momentum~constraint \nonumber
\end{eqnarray}

Similarly, we can expand (\ref{AA}) upto fourth order by expanding the
field variables ($\varphi_2, \phi_2, B_2$) and vary the action with
respect to second order perturbed field variables to obtain second
order equations. Fourth order action containing only $\varphi_2$ terms
are, {\scriptsize
\begin{equation}
\label{AP20}
\begin{split}
^{(4)}\mathcal{S}^{NC}_{\varphi_2} =&\int d^4x \Big(\frac{1}{4} P_X \phi_2
\varphi_0^{\prime} \varphi_2^{\prime} a^{2} + \frac{1}{4} P_X
{\delta}^{i j} \varphi_0^{\prime} \varphi_2^{\prime}
{\partial}_{i}{B_1} {\partial}_{j}{B_1} a^{2} - \frac{3}{4} P_X
\varphi_0^{\prime} \varphi_2^{\prime} \phi_1^{2} a^{2} + \frac{1}{2}
P_X \phi_1 \varphi_1^{\prime} \varphi_2^{\prime} a^{2} - \frac{1}{8}
P_X {\varphi_2^{\prime}}^{2} a^{2} +\\
&
 \frac{1}{4} P_X {\delta}^{i j}
\varphi_0^{\prime} {\partial}_{i}{B_2} {\partial}_{j}{\varphi_2} a^{2}
- \frac{1}{2} P_X \phi_1 {\delta}^{i j} \varphi_0^{\prime}
{\partial}_{i}{B_1} {\partial}_{j}{\varphi_2} a^{2} + \frac{1}{2} P_X
{\delta}^{i j} \varphi_1^{\prime} {\partial}_{i}{B_1}
{\partial}_{j}{\varphi_2} a^{2} + \frac{1}{2} P_X {\delta}^{i j}
\varphi_2^{\prime} {\partial}_{i}{B_1} {\partial}_{j}{\varphi_1} a^{2}
+ \\
&
\frac{1}{8} P_X {\delta}^{i j} {\partial}_{i}{\varphi_2}
{\partial}_{j}{\varphi_2} a^{2} + 
\frac{3}{2} P_{XX}
\varphi_2^{\prime} \phi_1^{2} {\varphi_0^{\prime}} ^{3} - 2 P_{XX}
\phi_1 \varphi_1^{\prime} \varphi_2^{\prime} {\varphi_0^{\prime}}^{2}
+ \frac{1}{2} P_{XX} \phi_1 {\delta}^{i j} {\partial}_{i}{B_1}
{\partial}_{j}{\varphi_2} {\varphi_0^{\prime}}^{3} + \frac{1}{2}
P_{XX} \phi_1 {\delta}^{i j} {\partial}_{i}{\varphi_1}
{\partial}_{j}{\varphi_2} {\varphi_0^{\prime}}^{2} -\\
&
 \frac{1}{4}
P_{XX} \phi_2 \varphi_2^{\prime} {\varphi_0^{\prime}} ^{3} -
\frac{1}{4} P_{XX} {\delta}^{i j} \varphi_2^{\prime}
{\partial}_{i}{B_1} {\partial}_{j}{B_1} {\varphi_0^{\prime}} ^{3} +
\frac{3}{4} P_{XX} \varphi_0^{\prime} \varphi_2^{\prime}
{\varphi_1^{\prime}} ^{2} - \frac{1}{2} P_{XX} {\delta}^{i j}
\varphi_1^{\prime} {\partial}_{i}{B_1} {\partial}_{j}{\varphi_2}
{\varphi_0^{\prime}} ^{2} - \\
&
\frac{1}{2} P_{XX} {\delta}^{i j}
\varphi_0^{\prime} \varphi_1^{\prime} {\partial}_{i}{\varphi_1}
{\partial}_{j}{\varphi_2} + \frac{1}{8} P_{XX} {\varphi_0^{\prime}}
^{2} {\varphi_2^{\prime} }^{2} - \frac{1}{2} P_{XX} {\delta}^{i j}
\varphi_2^{\prime} {\partial}_{i}{B_1} {\partial}_{j}{\varphi_1}
{\varphi_0^{\prime} }^{2} - \frac{1}{4} P_{XX} {\delta}^{i j}
\varphi_0^{\prime} \varphi_2^{\prime} {\partial}_{i}{\varphi_1}
{\partial}_{j}{\varphi_1} + \\
&
\frac{1}{8} P_{\varphi \varphi}
\varphi_2^{2} a^{4} + \frac{1}{4} P_{X\varphi} \phi_2
{\varphi_0^{\prime} }^{2} a^{2} \varphi_2 + \frac{1}{4} P_{X\varphi}
{\delta}^{i j} {\partial}_{i}{B_1} {\partial}_{j}{B_1}
{\varphi_0^{\prime} }^{2} a^{2} \varphi_2 - \frac{1}{2} P_{X\varphi}
\phi_1^{2} {\varphi_0^{\prime} }^{2} a^{2} \varphi_2 + \frac{1}{2}
P_{X\varphi} \phi_1 \varphi_0^{\prime} \varphi_1^{\prime} a^{2}
\varphi_2 + \\
&
\frac{1}{2} P_{X\varphi} \phi_1 \varphi_0^{\prime}
\varphi_2^{\prime} a^{2} \varphi_1 - \frac{1}{4} P_{X\varphi}
\varphi_0^{\prime} \varphi_2^{\prime} a^{2} \varphi_2 - \frac{1}{4}
P_{X\varphi} {\varphi_1^{\prime} }^{2} a^{2} \varphi_2 - \frac{1}{2}
P_{X\varphi} \varphi_1^{\prime} \varphi_2^{\prime} a^{2} \varphi_1 +
\frac{1}{2} P_{X\varphi} {\delta}^{i j} \varphi_0^{\prime}
{\partial}_{i}{B_1} {\partial}_{j}{\varphi_1} a^{2} \varphi_2 +\\
&
\frac{1}{2} P_{X\varphi} {\delta}^{i j} \varphi_0^{\prime}
{\partial}_{i}{B_1} {\partial}_{j}{\varphi_2} a^{2} \varphi_1 +
\frac{1}{4} P_{X\varphi} {\delta}^{i j} {\partial}_{i}{\varphi_1}
{\partial}_{j}{\varphi_1} a^{2} \varphi_2 + \frac{1}{2} P_{X\varphi}
{\delta}^{i j} {\partial}_{i}{\varphi_1} {\partial}_{j}{\varphi_2}
a^{2} \varphi_1 - \frac{1}{4} P_{XXX} \varphi_2^{\prime} \phi_1^{2}
{\varphi_0^{\prime} }^{5} a^{-2} + \\
&
\frac{1}{2} P_{XXX} \phi_1
\varphi_1^{\prime} \varphi_2^{\prime} {\varphi_0^{\prime} }^{4} a^{-2}
- \frac{1}{4} P_{XXX} \varphi_2^{\prime} {\varphi_0^{\prime} }^{3}
{\varphi_1^{\prime} }^{2} a^{-2} + \frac{1}{4} P_{\varphi \varphi
  \varphi} \varphi_1^{2} a^{4} \varphi_2 + \frac{1}{4} P_{XX\varphi}
\phi_1^{2} {\varphi_0^{\prime} }^{4} \varphi_2 - \frac{1}{2}
P_{XX\varphi} \phi_1 \varphi_1^{\prime} {\varphi_0^{\prime} }^{3}
\varphi_2 - \\
&
\frac{1}{2} P_{XX\varphi} \phi_1 \varphi_2^{\prime}
{\varphi_0^{\prime} }^{3} \varphi_1 + \frac{1}{4} P_{XX\varphi}
{\varphi_0^{\prime}}^{2} {\varphi_1^{\prime} }^{2} \varphi_2 +
\frac{1}{2} P_{XX\varphi} \varphi_1^{\prime} \varphi_2^{\prime}
{\varphi_0^{\prime} }^{2} \varphi_1 + \frac{1}{2} P_{X\varphi \varphi}
\phi_1 {\varphi_0^{\prime} }^{2} a^{2} \varphi_1 \varphi_2 -
\frac{1}{2} P_{X\varphi \varphi} \varphi_0^{\prime} \varphi_1^{\prime}
a^{2} \varphi_1 \varphi_2 - \\
&
\frac{1}{4} P_{X\varphi \varphi}
\varphi_0^{\prime} \varphi_2^{\prime} \varphi_1^{2} a^{2} +
\frac{1}{2} P_X \phi_1 {\delta}^{i j} {\partial}_{i}{\varphi_1}
{\partial}_{j}{\varphi_2} a^{2} + \frac{1}{2} P_{\varphi \varphi}
\phi_1 a^{4} \varphi_1 \varphi_2 + \frac{1}{4} P_\varphi \phi_2 a^{4}
\varphi_2 + \frac{1}{4} P_\varphi {\delta}^{i j} {\partial}_{i}{B_1}
{\partial}_{j}{B_1} a^{4} \varphi_2 \\
&- \frac{1}{4} P_\varphi \phi_1^{2}
a^{4} \varphi_2 \Big)
\end{split}
\end{equation}}

Varying the action with respect to $\varphi_2$ leads to the same
equation of motion of $\varphi_2$ (\ref{eos2}). Similarly second order
equations of $\phi_2$ and $B_2$ can be obtained from varying fourth
order action with respect to $\phi_2$ and $B_2$ (\ref{G002}) and
(\ref{G0i2}), respectively. This mechanism can be extended up to any
order and we can generalize that, \emph{equations obtained from both
  the approaches are identical and there are no ambiguities as
  discussed in Ref.} \citep{Appignani2010a}.

Another way of seeing constraints is, action (\ref{AP1}) or
(\ref{AP20}) contain no time derivative of $\phi_1, B_1, \phi_2$ and
$B_2$, i.e., Lapse function and Shift vector algebraically enter in
the action. Hence, variation with respect to $\phi$ and $B$ always
lead to constraint equations. So, we can use (\ref{P1}) and (\ref{P2})
constraint equations to eliminate $\phi_1$ and $B_1$ from the action
and use background equations (\ref{Back1}) and (\ref{Back2}) to obtain
a second order single variable action in terms of
$\varphi_1$. Further, writing the action in terms of Mukhanov-Sasaki
variable `$v$',
\begin{equation}\label{1stnc}
  ^{(2)}\mathcal{S}^{NC} = \frac{1}{2} \int d^4x \Big\{ {v^\prime}^2 -
  c_s^2~ \delta^{i j}~\partial_{i}{v}\partial_{j}{v} + \frac{z^{\prime
      \prime}}{z} v^2 \Big\}
\end{equation}
where $v, z$ and $c_s$ are defined in equation (\ref{vzc}). We can
vary the action (\ref{1stnc}) with respect to $v$ to obtain equation
of motion of $v$, which is identical to the equation (\ref{S1}).
Hence \emph{at first order, order-by-order Einstein's equation
  approach and reduced action approach lead to identical result.}

Similar procedure may be followed to obtain reduced second order
single variable equation of motion. Fourth order action does not
contain terms that have time derivatives of $\phi_2$ and $B_2$. Hence,
as in the first order, one should be able to substitute $\phi_2$ and
$B_2$ in the fourth order action to obtain a reduced action in terms
of $\varphi_2$. Malik et al showed, for canonical scalar field under
slow roll approximation, that the single variable equation from both
approaches are same\citep{Malik2008}. Similarly, since in the case of
non-canonical scalar field, equations of Lapse function $\phi_2$ and
Shift vector $\partial_i B_2$ are (1) identical for both the
approaches and (2) are constraint equations, reduced single variable
form of the equation of motion should also be identical.  In Appendix
\ref{App:Appendix B}, we give the reduced single variable action as
well as equation of motion in terms of `$\varphi_2$' for non-canonical
scalar field in Power law limit.  Hence, \emph{both approaches give
  the identical results up to second order.}

At the first order, equations of motion are linear in first order
variables. In higher order, only the highest order perturbed variables
appear linearly, where the lowest order perturbed variables contribute
non-linearly to equations of motion. For example, in second order,
equations are linear in second order variables $\varphi_2, \phi_2$ and
$B_2$ but are quadratic in first order variables $\varphi_1, \phi_1$
and $B_1$. Hence, as pointed out in \citep{Appignani2010a}, it does
appear that obtaining equations of field variables $\phi, B$ and
$\varphi$ at higher order from the two approaches may not be identical
and thus the reduced form of the single variable equations of motion
of the two different approaches may differ. However, \emph{instead of
  the non-linear form of the perturbed action, reduced single variable
  equations of motion at second order obtained from both the
  approaches are identical, hence we can generalize that at every
  order, in the case of non-canonical scalar field, both approaches
  lead to identical result.}  This leads to the following question:
\begin{center}
 Why Appignani et al\citep{Appignani2010a} obtained different
  equations of motion from two approaches?
\end{center}

In the simplified model proposed in Ref.\citep{Appignani2010a},
authors have assumed that the homogeneous universe is filled with
matter fluctuations with no Lapse function $\phi$ and Shift vector
$\partial_{i}{B}$. They have shown that in this simplified model
stress tensor, energy density and pressure are not identical for both
the approaches. Note that, since there are no metric fluctuations,
left hand sides of the perturbed Einstein's equations are zero. This
leads to five equations ($T^{0}_{0} = 0, T^{0}_{i} = 0$ and equation
of motion of scalar field) for a single perturbed variable
$\delta\varphi$, which lead to the inconsistency of the simplified
model, hence the ambiguities. Another way of looking into this is the
following: in the action approach, one can obtain the Hamiltonian
(momentum) constraint of the system by varying the action with respect
to $\phi$ ($B$). Since in this simplified model, both are not present,
this leads to inconsistent results.

This leads to another important question which we address in the rest
of the paper is:
\begin{center}
  For what theories of gravity and matter field, the two approaches
  lead to identical single variable equation of motion?
\end{center} 

To answer the question, let us look at the procedure of conventional
gauge invariant cosmological perturbation theory, which is based on
two things, first, to obtain gauge invariant variables and second, to
obtain a single variable action/equation of motion in terms of gauge
invariant variables. Gauge invariant variables are model independent
(if the background metric is unchanged), i.e., these are same for
canonical, non-canonical or Galilean models so that we can always
remove two variables out of five by using gauge conditions and define
suitable gauge invariant variables. At each order, we start from five
perturbed variables ($\phi, B, \psi, E$ and $\varphi$). The gauge
choice helps to remove two variables. Carefully choosing a gauge (in
our case, $E=0$ and $\psi=0$) at any order could fix the gauge issue
and reduced variables will coincide with gauge invariant
variables\citep{Malik2009}. So all equations in terms of those
variables also become gauge invariant.

Obtaining a single variable action/equation of motion depends solely
on gauge fixing (the procedure discussed in the above paragraph) and
two constraint equations which differ from model to model. If Lapse
function $N$ ($\phi$ in perturbed case) and Shift vector $N_i$
($\partial_i B$ in perturbed case) remain constraints for any models,
i.e., those functions algebraically enter into the action then
equations of motion of Lapse function and Shift vector contain no time
derivatives of them, we can always eliminate them from action/equation
of motion to get a single variable action/equation. This helps to
reduce the degrees of freedom to one and we can write the
action/equation of motion in a single variable form. However, if
$\phi_1$ or $B_1$ or both become dynamical i.e., if the action
contains terms containing time derivatives of Lapse function and/or
Shift vector such that equations contain double time derivatives of
those variables then it is not possible to substitute those variables
in the action or in the equation of motion of the scalar field and the
method fails. We refer the constrained nature of Lapse function and
Shift vector as `Constraint consistency' condition. If it is satisfied
then the whole method of gauge invariant cosmological perturbation
theory will work. In the next section, we test the `constraint
consistency' condition for several models that are used in the
literature.

In fact, the whole exercise may be done in terms of Lapse $N$, Shift
$N_{i}$ and scalar field $\varphi$ without applying any perturbation
theory. From Hamiltonian theory of General relativity or from
Einstein's equation we obtain constraint equations, i.e., Hamiltonian
and Momentum constraints which are functions of $(N, N_{i}, \gamma_{i
  j}, \varphi, \varphi^\prime, \partial_{i}\varphi)$ in which Lapse
function and Shift vector are constraints. If out of four constraint
equations (1 Hamiltonian equation or 0-0 Einstein's equation and 3
Momentum constraint equation or 0-i Einstein's equation), we can solve
and extract four quantity, one Lapse function and three component
Shift vector then we can substitute those back in the action or in
equation of motion of the scalar field. Unfortunately, GR equations
are so highly non-linear that analytically solving constraint
equations for Lapse function and Shift vector and obtaining a single
variable action or equation is very difficult. Perturbation theory
helps to simplify those equations so that we can invert those
equations in terms of Lapse function and Shift vector and obtain a
simplified solutions of constraint functions.

\section{Specific models}\label{sec3}
In this section, first we start with well known models of inflation
within the framework of general relativity and then move to modified
gravity models. To check for constraint consistency we follow action
formulation, write down second order action in terms of perturbed
variables and identify terms that contain time derivatives of Lapse
function and/or Shift vector.

\subsection{Minimally coupled Galilean field}
We start with the model with derivatives of metric in the action, 
\begin{equation}
 \mathcal{S}^G_m = \int d^4 x \sqrt{-g}~ G(X, \varphi) \Box \varphi 
\end{equation}
which has been proposed by Kobayashi et
al\citep{Kobayashi2010,Kobayashi2011} where `$\Box$' is defined by
equation (\ref{boxx}). Since `$\Box$' contains time derivatives of
metric as well as matter, it is not obvious whether the action can be
expressed in a single variable form. After partial integration, the
second order matter action becomes as follows:

{\scriptsize
\begin{equation}
\begin{split}
  ^{(2)}\mathcal{S}^G_{m} =& \int d^4x \Big\{ 5 \bm{{G_X \phi_1
      \phi_1^\prime} {\varphi_0^\prime }^{3}} + \frac{15}{2} G_X
  \varphi_0^{\prime \prime} \phi_1^{2} {\varphi_0^\prime }^{2} -
  \frac{15}{2} G_X a^\prime \phi_1^{2} {\varphi_0^\prime }^{3} a^{-1}
  - \bm{3 G_X \phi_1^\prime \varphi_1^\prime {\varphi_0^\prime }^{2}}
  - 6 G_X \phi_1 \varphi_0^\prime {\varphi_1^\prime} \varphi_0^{\prime
    \prime} + \\
    &
    9 G_X \phi_1 \varphi_1^\prime a^\prime {\varphi_0^\prime
  }^{2} a^{-1} - G_X {\delta}^{i j} {\partial}_{i}{B_1} {\partial}_{0
    j}{B_1} {\varphi_0^\prime }^{3} - \frac{3}{2} G_X {\delta}^{i j}
  {\partial}_{i}{B_1} {\partial}_{j}{B_1} \varphi_0^{\prime \prime}
  {\varphi_0^\prime }^{2} + \frac{3}{2} G_X {\delta}^{i j} a^\prime
  {\partial}_{i}{B_1} {\partial}_{j}{B_1} {\varphi_0^\prime }^{3}
  a^{-1} - \\
  &
  3 G_X \phi_1 \varphi_1^{\prime \prime} {\varphi_0^\prime
  }^{2} + G_X \varphi_0^{\prime \prime} {\varphi_1^\prime }^{2} + 2
  G_X \varphi_0^\prime \varphi_1^\prime \varphi_1^{\prime \prime} - 3
  G_X \varphi_0^\prime a^\prime {\varphi_1^\prime }^{2} a^{-1} - 
  2 G_X
  {\delta}^{i j} \varphi_0^\prime {\partial}_{i}{B_1}
  {\partial}_{j}{\varphi_1} \varphi_0^{\prime \prime} - \\
  &
  G_X
  {\delta}^{i j} {\partial}_{i}{\varphi_1} {\partial}_{0 j}{B_1}
  {\varphi_0^\prime }^{2} - 2 G_X {\delta}^{i j} {\partial}_{i}{B_1}
  {\partial}_{0 j}{\varphi_1} {\varphi_0^\prime }^{2} + 3 G_X
  {\delta}^{i j} a^\prime {\partial}_{i}{B_1}
  {\partial}_{j}{\varphi_1} {\varphi_0^\prime }^{2} a^{-1} - 
  2 G_X
  {\delta}^{i j} \varphi_0^\prime {\partial}_{i}{\varphi_1}
  {\partial}_{0 j}{\varphi_1} + \\
  &
  G_X {\delta}^{i j} \varphi_0^\prime
  a^\prime {\partial}_{i}{\varphi_1} {\partial}_{j}{\varphi_1} a^{-1}
  + G_X {\delta}^{i j} {\partial}_{i}{B_1} {\partial}_{j}{\phi_1}
  {\varphi_0^\prime }^{3} + G_X {\delta}^{i j} {\partial}_{i}{\phi_1}
  {\partial}_{j}{\varphi_1} {\varphi_0^\prime }^{2} - \bm{G_{XX}
    \phi_1 \phi_1^\prime {\varphi_0^\prime }^{5} a^{(-2)}} - \\
    &
    5 G_{XX}
  \varphi_0^{\prime \prime} \phi_1^{2} {\varphi_0^\prime }^{4}
  a^{(-2)} + 5 G_{XX} a^\prime \phi_1^{2} {\varphi_0^\prime }^{5}
  a^{(-3)} + 7 G_{XX} \phi_1 \varphi_1^\prime \varphi_0^{\prime
    \prime} {\varphi_0^\prime }^{3} a^{(-2)} - 8 G_{XX} \phi_1
  \varphi_1^\prime a^\prime {\varphi_0^\prime }^{4} a^{(-3)} +\\
  &
   G_{XX}
  \phi_1 \varphi_1^{\prime \prime} {\varphi_0^\prime }^{4} a^{(-2)} +
  \frac{1}{2} G_{XX} {\delta}^{i j} {\partial}_{i}{B_1}
  {\partial}_{j}{B_1} \varphi_0^{\prime \prime} {\varphi_0^\prime
  }^{4} a^{(-2)} - \frac{1}{2} G_{XX} {\delta}^{i j} a^\prime
  {\partial}_{i}{B_1} {\partial}_{j}{B_1} {\varphi_0^\prime }^{5}
  a^{(-3)} + \bm{G_{XX} \phi_1^\prime \varphi_1^\prime
    {\varphi_0^\prime }^{4} a^{(-2)}}\\
    &
     - \frac{5}{2} G_{XX}
  \varphi_0^{\prime \prime} {\varphi_0^\prime }^{2} {\varphi_1^\prime
  }^{2} a^{(-2)} + \frac{7}{2} G_{XX} a^\prime {\varphi_0^\prime }^{3}
  {\varphi_1^\prime }^{2} a^{(-3)} - G_{XX} \varphi_1^\prime
  \varphi_1^{\prime \prime} {\varphi_0^\prime }^{3} a^{(-2)} + G_{XX}
  {\delta}^{i j} {\partial}_{i}{B_1} {\partial}_{j}{\varphi_1}
  \varphi_0^{\prime \prime} {\varphi_0^\prime }^{3} a^{(-2)} - \\
  &
  G_{XX}
  {\delta}^{i j} a^\prime {\partial}_{i}{B_1}
  {\partial}_{j}{\varphi_1} {\varphi_0^\prime }^{4} a^{(-3)} +
  \frac{1}{2} G_{XX} {\delta}^{i j} {\partial}_{i}{\varphi_1}
  {\partial}_{j}{\varphi_1} \varphi_0^{\prime \prime}
  {\varphi_0^\prime }^{2} a^{(-2)} - \frac{1}{2} G_{XX} {\delta}^{i j}
  a^\prime {\partial}_{i}{\varphi_1} {\partial}_{j}{\varphi_1}
  {\varphi_0^\prime }^{3} a^{(-3)} -\\
  &
   \bm{G_{XY} \phi_1^\prime
    {\varphi_0^\prime }^{3} \varphi_1} - 3 G_{XY} \phi_1
  \varphi_0^{\prime \prime} {\varphi_0^\prime }^{2} \varphi_1 + 3
  G_{XY} \phi_1 a^\prime {\varphi_0^\prime }^{3} a^{-1} \varphi_1 + 2
  G_{XY} \varphi_0^\prime \varphi_1^\prime \varphi_0^{\prime \prime}
  \varphi_1 - 3 G_{XY} \varphi_1^\prime a^\prime {\varphi_0^\prime
  }^{2} a^{-1} \varphi_1 + \\
  &
  G_{XY} \varphi_1^{\prime \prime}
  {\varphi_0^\prime }^{2} \varphi_1 + \frac{1}{2} G_{XXX}
  \varphi_0^{\prime \prime} \phi_1^{2} {\varphi_0^\prime }^{6}
  a^{(-4)} - \frac{1}{2} G_{XXX} a^\prime \phi_1^{2} {\varphi_0^\prime
  }^{7} a^{(-5)} - G_{XXX} \phi_1 \varphi_1^\prime \varphi_0^{\prime
    \prime} {\varphi_0^\prime }^{5} a^{(-4)} + \\
    &
    G_{XXX} \phi_1
  \varphi_1^\prime a^\prime {\varphi_0^\prime }^{6} a^{(-5)} +
  \frac{1}{2} G_{XXX} \varphi_0^{\prime \prime} {\varphi_0^\prime
  }^{4} {\varphi_1^\prime }^{2} a^{(-4)} - \frac{1}{2} G_{XXX}
  a^\prime {\varphi_0^\prime }^{5} {\varphi_1^\prime }^{2} a^{(-5)} +
  \frac{1}{2} G_{XYY} \varphi_0^{\prime \prime} {\varphi_0^\prime
  }^{2} \varphi_1^{2} - \\
  &
  \frac{1}{2} G_{XYY} a^\prime {\varphi_0^\prime
  }^{3} \varphi_1^{2} a^{-1} + G_{XXY} \phi_1 \varphi_0^{\prime
    \prime} {\varphi_0^\prime }^{4} a^{(-2)} \varphi_1 - G_{XXY}
  \phi_1 a^\prime {\varphi_0^\prime }^{5} a^{(-3)} \varphi_1 - G_{XXY}
  \varphi_1^\prime \varphi_0^{\prime \prime} {\varphi_0^\prime }^{3}
  a^{(-2)} \varphi_1 + \\
  &
  G_{XXY} \varphi_1^\prime a^\prime
  {\varphi_0^\prime }^{4} a^{(-3)} \varphi_1 + \frac{1}{4} G_Y
  {\delta}^{i j} {\partial}_{i}{B_1} {\partial}_{j}{B_1}
  {\varphi_0^\prime }^{2} a^{2} - \frac{3}{4} G_Y \phi_1^{2}
  {\varphi_0^\prime }^{2} a^{2} + G_Y \phi_1 \varphi_0^\prime
  \varphi_1^\prime a^{2} - \frac{1}{2} G_Y {\varphi_1^\prime }^{2}
  a^{2}
  + \\
  &
  G_Y {\delta}^{i j} \varphi_0^\prime {\partial}_{i}{B_1}
  {\partial}_{j}{\varphi_1} a^{2} + \frac{1}{2} G_Y {\delta}^{i j}
  {\partial}_{i}{\varphi_1} {\partial}_{j}{\varphi_1} a^{2} +
  \frac{3}{2} G_{XY} \phi_1^{2} {\varphi_0^\prime }^{4} - \frac{5}{2}
  G_{XY} \phi_1 \varphi_1^\prime {\varphi_0^\prime }^{3} - \frac{1}{4}
  G_{XY} {\delta}^{i j} {\partial}_{i}{B_1} {\partial}_{j}{B_1}
  {\varphi_0^\prime }^{4} + \\
  &
  \frac{5}{4} G_{XY} {\varphi_0^\prime }^{2}
  {\varphi_1^\prime }^{2} - \frac{1}{2} G_{XY} {\delta}^{i j}
  {\partial}_{i}{B_1} {\partial}_{j}{\varphi_1} {\varphi_0^\prime
  }^{3} - \frac{1}{4} G_{XY} {\delta}^{i j} {\partial}_{i}{\varphi_1}
  {\partial}_{j}{\varphi_1} {\varphi_0^\prime }^{2} + \frac{1}{2}
  G_{YY} \phi_1 {\varphi_0^\prime }^{2} a^{2} \varphi_1 - G_{YY}
  \varphi_0^\prime \varphi_1^\prime a^{2} \varphi_1 \\
  &
  - \frac{1}{4}
  G_{XXY} \phi_1^{2} {\varphi_0^\prime }^{6} a^{(-2)} + \frac{1}{2}
  G_{XXY} \phi_1 \varphi_1^\prime {\varphi_0^\prime }^{5} a^{(-2)} -
  \frac{1}{4} G_{XXY} {\varphi_0^\prime }^{4} {\varphi_1^\prime }^{2}
  a^{(-2)} - \frac{1}{4} G_{YYY} {\varphi_0^\prime }^{2} \varphi_1^{2}
  a^{2} -\\
  &
   \frac{1}{2} G_{XYY} \phi_1 {\varphi_0^\prime }^{4} \varphi_1
  + \frac{1}{2} G_{XYY} \varphi_1^\prime {\varphi_0^\prime }^{3}
  \varphi_1 a^{2} \Big\}
\end{split}
\end{equation}
  }

  \noindent where $G_X \equiv \partial_{X}{G}, G_Y
  \equiv \partial_{\varphi}{G}$ for background and so on. The
  derivatives of the constraints ($\phi_1$) appear linearly in the
  above reduced action (those terms are highlighted in the above
  expression). However, by performing partial integration one can
  rewrite these terms as terms proportional to $\phi_1$ e.g., first
  terms in the action can be written as $ -
  \frac{5}{2} \partial_{0}\{G_X {\varphi_0^\prime}^3\} \phi_1^2$. So,
  \textit{although the action contains time derivative of Lapse
    function but it is reducible to action with no time derivative of
    Lapse or Shift, hence, variation of these terms do not lead
    constraint inconsistencies.}

\subsection{\texorpdfstring{$({\varphi}_{;\lambda} {\varphi}^{; \lambda} \{\Box
  \varphi\}^2)$}{TEXT} model}
Let us consider the following model where the matter action is given
by

\begin{equation}
\mathcal{S}_m = \int d^4x \sqrt{-g} {\varphi}_{;\lambda} {\varphi}^{;\lambda} 
\{\Box \varphi\}^2
\end{equation}
and is minimally coupled to gravity\footnote{Note that, our motivation
  is only about the consistency of the method discussed in the first
  section for different models, not to check its physical
  observational viability or any other problems such as Higher order
  Ostrogradsky's ghost}. Expanding the matter action to second order,
we get,

{\scriptsize
\begin{equation}
\label{box}
\begin{split}
^{(2)}\mathcal{S}_m =& \int d^4x \{- \frac{35}{2} \phi_1^{2}
{\varphi_0^{\prime}} ^{2} {\varphi_0^{\prime \prime} }^{2} a^{-2} - 70
a^\prime \varphi_0^{\prime \prime} \phi_1^{2} {\varphi_0^{\prime}
}^{3} a^{-3} - 14 \phi_1 \phi_1^{\prime} \varphi_0^{\prime \prime}
{\varphi_0^{\prime} }^{3} a^{-2} - 10 \phi_1 {\delta}^{i j}
\varphi_0^{\prime \prime} {\partial}_{i j}{B_1} {\varphi_0^{\prime}
}^{3} a^{-2} - \\
&
70 \phi_1^{2} {\varphi_0^{\prime} }^{4} {a^\prime }^{2}
a^{-4} - 28 \phi_1 \phi_1^{\prime} a^\prime {\varphi_0^{\prime} }^{4}
a^{-3} - 20 \phi_1 {\delta}^{i j} a^\prime {\partial}_{i j}{B_1}
{\varphi_0^{\prime} }^{4} a^{-3} + 10 \phi_1 \varphi_0^{\prime \prime}
\varphi_1^{\prime \prime} {\varphi_0^{\prime} }^{2} a^{-2} - 6 \phi_1
{\delta}^{i j} \varphi_0^{\prime \prime} {\partial}_{j i}{\varphi_1}
{\varphi_0^{\prime} }^{2} a^{-2} \\
&
+ 20 \phi_1 a^\prime
\varphi_1^{\prime \prime} {\varphi_0^{\prime} }^{3} a^{-3} + 60 \phi_1
\varphi_1^{\prime} a^\prime \varphi_0^{\prime \prime}
{\varphi_0^{\prime} }^{2} a^{-3} - 12 \phi_1 {\delta}^{i j} a^\prime
{\partial}_{j i}{\varphi_1} {\varphi_0^{\prime} }^{3} a^{-3} + 80
\phi_1 \varphi_1^{\prime} {\varphi_0^{\prime} }^{3} {a^\prime }^{2}
a^{-4} + 10 \phi_1 \varphi_0^{\prime} \varphi_1^{\prime}
{\varphi_0^{\prime \prime} }^{2} a^{-2}\\
&
 + 6 \phi_1^{\prime}
\varphi_1^{\prime} \varphi_0^{\prime \prime} {\varphi_0^{\prime} }^{2}
a^{-2} + 6 {\delta}^{i j} \varphi_1^{\prime} \varphi_0^{\prime \prime}
{\partial}_{i j}{B_1} {\varphi_0^{\prime} }^{2} a^{-2} + 16
\phi_1^{\prime} \varphi_1^{\prime} a^\prime {\varphi_0^{\prime} }^{3}
a^{-3} + 16 {\delta}^{i j} \varphi_1^{\prime} a^\prime {\partial}_{i
  j}{B_1} {\varphi_0^{\prime} }^{3} a^{-3} - 4 \varphi_0^{\prime}
\varphi_1^{\prime} \varphi_0^{\prime \prime} \varphi_1^{\prime \prime}
a^{-2}
\\
&
+ 4 {\delta}^{i j} \varphi_0^{\prime} \varphi_1^{\prime}
\varphi_0^{\prime \prime} {\partial}_{j i}{\varphi_1} a^{-2} - 12
\varphi_1^{\prime} a^\prime \varphi_1^{\prime \prime}
{\varphi_0^{\prime} }^{2} a^{-3} - 12 \varphi_0^{\prime} a^\prime
\varphi_0^{\prime \prime} {\varphi_1^{\prime} }^{2} a^{-3} + 12
{\delta}^{i j} \varphi_1^{\prime} a^\prime {\partial}_{j i}{\varphi_1}
{\varphi_0^{\prime} }^{2} a^{-3} - 24 {\varphi_0^{\prime} }^{2}
{\varphi_1^{\prime} }^{2} {a^\prime }^{2} a^{-4} \\
&
+ \frac{5}{2}
{\delta}^{i j} {\partial}_{i}{B_1} {\partial}_{j}{B_1}
{\varphi_0^{\prime} }^{2} {\varphi_0^{\prime \prime} }^{2} a^{-2} + 10
{\delta}^{i j} a^\prime {\partial}_{i}{B_1} {\partial}_{j}{B_1}
\varphi_0^{\prime \prime} {\varphi_0^{\prime} }^{3} a^{-3} + 10
{\delta}^{i j} {\partial}_{i}{B_1} {\partial}_{j}{B_1}
{\varphi_0^{\prime} }^{4} {a^\prime }^{2} a^{-4} - {\varphi_1^{\prime}
}^{2} {\varphi_0^{\prime \prime} }^{2} a^{-2} + \\
&
2 {\delta}^{i j}
\varphi_0^{\prime} {\partial}_{j}{B_1} {\partial}_{i}{\varphi_1}
{\varphi_0^{\prime \prime} }^{2} a^{-2} + 8 {\delta}^{i j} a^\prime
{\partial}_{j}{B_1} {\partial}_{i}{\varphi_1} \varphi_0^{\prime
  \prime} {\varphi_0^{\prime} }^{2} a^{-3} + 8 {\delta}^{i j}
{\partial}_{j}{B_1} {\partial}_{i}{\varphi_1} {\varphi_0^{\prime}
}^{3} {a^\prime }^{2} a^{-4} + {\delta}^{i j}
{\partial}_{i}{\varphi_1} {\partial}_{j}{\varphi_1} {\varphi_0^{\prime
    \prime} }^{2} a^{-2} + \\
    &
    4 {\delta}^{i j} \varphi_0^{\prime}
a^\prime {\partial}_{i}{\varphi_1} {\partial}_{j}{\varphi_1}
\varphi_0^{\prime \prime} a^{-3} + 4 {\delta}^{i j}
{\partial}_{i}{\varphi_1} {\partial}_{j}{\varphi_1}
{\varphi_0^{\prime} }^{2} {a^\prime }^{2} a^{-4} + 4 {\delta}^{i j}
{\partial}_{i}{B_1} \varphi_0^{\prime \prime} {\partial}_{j
  0}{\varphi_1} {\varphi_0^{\prime} }^{2} a^{-2} + 2 \phi_1^{\prime}
\varphi_1^{\prime \prime} {\varphi_0^{\prime} }^{3} a^{-2} +\\
&
 4
{\delta}^{i j} a^\prime {\partial}_{i}{B_1} {\partial}_{j}{\varphi_1}
\varphi_0^{\prime \prime} {\varphi_0^{\prime} }^{2} a^{-3} + 2
{\delta}^{i j} {\partial}_{i}{\varphi_1} \varphi_0^{\prime \prime}
{\partial}_{0 j}{B_1} {\varphi_0^{\prime} }^{2} a^{-2} + 2 {\delta}^{i
  j} \varphi_1^{\prime \prime} {\partial}_{i j}{B_1}
{\varphi_0^{\prime} }^{3} a^{-2}
+ 8 {\delta}^{i j} a^\prime {\partial}_{i}{B_1} {\partial}_{j
  0}{\varphi_1} {\varphi_0^{\prime} }^{3} a^{-3} - \\
  &
  2 {\delta}^{i j}
\phi_1^{\prime} {\partial}_{j i}{\varphi_1} {\varphi_0^{\prime} }^{3}
a^{-2} - 2 {\delta}^{i j} {\delta}^{k l} {\partial}_{i j}{B_1}
{\partial}_{l k}{\varphi_1} {\varphi_0^{\prime} }^{3} a^{-2} + 2
{\delta}^{i j} {\partial}_{i}{\phi_1} {\partial}_{j}{\varphi_1}
\varphi_0^{\prime \prime} {\varphi_0^{\prime} }^{2} a^{-2} + 8
{\delta}^{i j} {\partial}_{i}{B_1} {\partial}_{j}{\varphi_1}
{\varphi_0^{\prime} }^{3} {a^\prime }^{2} a^{-4} + \\
&
4 {\delta}^{i j}
a^\prime {\partial}_{i}{\varphi_1} {\partial}_{0 j}{B_1}
{\varphi_0^{\prime} }^{3} a^{-3} + 4 {\delta}^{i j} a^\prime
{\partial}_{i}{\phi_1} {\partial}_{j}{\varphi_1} {\varphi_0^{\prime}
}^{3} a^{-3} + 2 {\delta}^{i j} {\partial}_{i}{B_1} \varphi_0^{\prime
  \prime} {\partial}_{0 j}{B_1} {\varphi_0^{\prime} }^{3} a^{-2} - 2
{\delta}^{i j} {\partial}_{i}{B_1} {\partial}_{j}{\phi_1}
\varphi_0^{\prime \prime} {\varphi_0^{\prime} }^{3} a^{-2} - \\
&
\bm{
  {\phi_1^\prime}^2 {\varphi_0^{\prime} }^{4} a^{-2}} + 4 {\delta}^{i
  j} a^\prime {\partial}_{i}{B_1} {\partial}_{0 j}{B_1}
{\varphi_0^{\prime} }^{4} a^{-3} - 4 {\delta}^{i j} a^\prime
{\partial}_{i}{B_1} {\partial}_{j}{\phi_1} {\varphi_0^{\prime} }^{4}
a^{-3} - 2 {\delta}^{i j} \phi_1^{\prime} {\partial}_{i j}{B_1}
{\varphi_0^{\prime} }^{4} a^{-2} - \\
&
{\delta}^{i j} {\delta}^{k l}
{\partial}_{i j}{B_1} {\partial}_{k l}{B_1} {\varphi_0^{\prime} }^{4}
a^{-2} - 
{\varphi_0^{\prime} }^{2} {\varphi_1^{\prime \prime} }^{2}
a^{-2} + 2 {\delta}^{i j} \varphi_1^{\prime \prime} {\partial}_{j
  i}{\varphi_1} {\varphi_0^{\prime} }^{2} a^{-2} - {\delta}^{i j}
{\delta}^{k l} {\partial}_{j i}{\varphi_1} {\partial}_{l k}{\varphi_1}
{\varphi_0^{\prime} }^{2} a^{-2} \}
\end{split}
\end{equation}}

Following points are worth-noting regarding the above action: (i)
Unlike Galilean scalar fields, the above action contains square of the
derivative of constraints (${\phi^\prime}^2$) [the term is highlighted
in the above expression], (ii) In case of Galilean, it was possible to
rewrite it as boundary term, in this case it is not possible. This
implies that using reduced action approach, we cannot obtain a single
dynamical equation. Hence as discussed in section (2), the constraint
consistency is not satisfied.

\subsection{\texorpdfstring{$(\varphi_{;\lambda} \varphi^{;\lambda}~ \varphi_{; \mu
    \nu} \varphi^{; \mu \nu})$}{TEXT} model}
As like the previous subsection, the following action also contains
higher time derivatives of constraints. Expanding the matter action
\begin{equation}
  \mathcal{S}_m = \int d^4x~\sqrt{-g}~ {\varphi}_{;\lambda} 
{\varphi}^{;\lambda}~ \varphi_{; \mu \nu} \varphi^{; \mu \nu}
\end{equation}
to second order, we get,

{\scriptsize
\begin{equation}
\label{boxd}
\begin{split}
^{(2)}\mathcal{S}_m = &\int d^4x \{ - \frac{35}{2} \phi_1^{2}
{\varphi_0^{\prime} }^{2} {\varphi_0^{\prime \prime} }^{2} a^{-2} + 35
a^\prime \varphi_0^{\prime \prime} \phi_1^{2} {\varphi_0^{\prime}
}^{3} a^{-3} - 14 \phi_1 \phi_1^{\prime} \varphi_0^{\prime \prime}
{\varphi_0^{\prime} }^{3} a^{-2} - 70 \phi_1^{2} {\varphi_0^{\prime}
}^{4} {a^\prime }^{2} a^{-4} + 14 \phi_1 \phi_1^{\prime} a^\prime
{\varphi_0^{\prime} }^{4} a^{-3} - \\
& 10 \phi_1 {\delta}^{i j} a^\prime
{\partial}_{i j}{B_1} {\varphi_0^{\prime} }^{4} a^{-3} + 10 \phi_1
\varphi_0^{\prime \prime} \varphi_1^{\prime \prime}
{\varphi_0^{\prime} }^{2} a^{-2} - 10 \phi_1 a^\prime
\varphi_1^{\prime \prime} {\varphi_0^{\prime} }^{3} a^{-3} - 30 \phi_1
\varphi_1^{\prime} a^\prime \varphi_0^{\prime \prime}
{\varphi_0^{\prime} }^{2} a^{-3} - 6 \phi_1 {\delta}^{i j} a^\prime
{\partial}_{j i}{\varphi_1} {\varphi_0^{\prime} }^{3} a^{-3} \\
& + 80\phi_1 \varphi_1^{\prime} {\varphi_0^{\prime} }^{3} {a^\prime }^{2}
a^{-4} + 10 \phi_1 \varphi_0^{\prime} \varphi_1^{\prime}
{\varphi_0^{\prime \prime} }^{2} a^{-2} + 6 \phi_1^{\prime}
\varphi_1^{\prime} \varphi_0^{\prime \prime} {\varphi_0^{\prime} }^{2}
a^{-2} - 8 \phi_1^{\prime} \varphi_1^{\prime} a^\prime
{\varphi_0^{\prime} }^{3} a^{-3} + 8 {\delta}^{i j} \varphi_1^{\prime}
a^\prime {\partial}_{i j}{B_1} {\varphi_0^{\prime} }^{3} a^{-3} - \\
& 4\varphi_0^{\prime} \varphi_1^{\prime} \varphi_0^{\prime \prime}
\varphi_1^{\prime \prime} a^{-2} + 6 \varphi_1^{\prime} a^\prime
\varphi_1^{\prime \prime} {\varphi_0^{\prime} }^{2} a^{-3} + 6
\varphi_0^{\prime} a^\prime \varphi_0^{\prime \prime}
{\varphi_1^{\prime} }^{2} a^{-3} + 6 {\delta}^{i j} \varphi_1^{\prime}
a^\prime {\partial}_{j i}{\varphi_1} {\varphi_0^{\prime} }^{2} a^{-3}
- 24 {\varphi_0^{\prime} }^{2} {\varphi_1^{\prime} }^{2} {a^\prime
}^{2} a^{-4} + \\
&\frac{5}{2} {\delta}^{i j} {\partial}_{i}{B_1}
{\partial}_{j}{B_1} {\varphi_0^{\prime} }^{2} {\varphi_0^{\prime
    \prime} }^{2} a^{-2} - 5 {\delta}^{i j} a^\prime
{\partial}_{i}{B_1} {\partial}_{j}{B_1} \varphi_0^{\prime \prime}
{\varphi_0^{\prime} }^{3} a^{-3} + 10 {\delta}^{i j}
{\partial}_{i}{B_1} {\partial}_{j}{B_1} {\varphi_0^{\prime} }^{4}
{a^\prime }^{2} a^{-4} - {\varphi_1^{\prime} }^{2} {\varphi_0^{\prime
    \prime} }^{2} a^{-2} + \\
    &
    2 {\delta}^{i j} \varphi_0^{\prime}
{\partial}_{j}{B_1} {\partial}_{i}{\varphi_1} {\varphi_0^{\prime
    \prime} }^{2} a^{-2} - 4 {\delta}^{i j} a^\prime
{\partial}_{j}{B_1} {\partial}_{i}{\varphi_1} \varphi_0^{\prime
  \prime} {\varphi_0^{\prime} }^{2} a^{-3} + 8 {\delta}^{i j}
{\partial}_{j}{B_1} {\partial}_{i}{\varphi_1} {\varphi_0^{\prime}
}^{3} {a^\prime }^{2} a^{-4} + {\delta}^{i j}
{\partial}_{i}{\varphi_1} {\partial}_{j}{\varphi_1} {\varphi_0^{\prime
    \prime} }^{2} a^{-2} - \\
    &
    2 {\delta}^{i j} \varphi_0^{\prime}
a^\prime {\partial}_{i}{\varphi_1} {\partial}_{j}{\varphi_1}
\varphi_0^{\prime \prime} a^{-3} + 6 {\delta}^{i j}
{\partial}_{i}{\varphi_1} {\partial}_{j}{\varphi_1}
{\varphi_0^{\prime} }^{2} {a^\prime }^{2} a^{-4} + 4 {\delta}^{i j}
{\partial}_{i}{B_1} \varphi_0^{\prime \prime} {\partial}_{j
  0}{\varphi_1} {\varphi_0^{\prime} }^{2} a^{-2} + 2 \phi_1^{\prime}
\varphi_1^{\prime \prime} {\varphi_0^{\prime} }^{3} a^{-2} - \\
&
2{\delta}^{i j} a^\prime {\partial}_{i}{B_1} {\partial}_{j}{\varphi_1}
\varphi_0^{\prime \prime} {\varphi_0^{\prime} }^{2} a^{-3} + 2
{\delta}^{i j} {\partial}_{i}{\varphi_1} \varphi_0^{\prime \prime}
{\partial}_{0 j}{B_1} {\varphi_0^{\prime} }^{2} a^{-2} - 4 {\delta}^{i
  j} a^\prime {\partial}_{i}{B_1} {\partial}_{j 0}{\varphi_1}
{\varphi_0^{\prime} }^{3} a^{-3} - 4 {\delta}^{i j}
{\partial}_{i}{\phi_1} {\partial}_{j 0}{\varphi_1} {\varphi_0^{\prime}
}^{3} a^{-2} - \\
&
2 {\delta}^{i j} {\delta}^{k l} {\partial}_{i k}{B_1}
{\partial}_{l j}{\varphi_1} {\varphi_0^{\prime} }^{3} a^{-2} + 2
{\delta}^{i j} {\partial}_{i}{\phi_1} {\partial}_{j}{\varphi_1}
\varphi_0^{\prime \prime} {\varphi_0^{\prime} }^{2} a^{-2} + 8
{\delta}^{i j} {\partial}_{i}{B_1} {\partial}_{j}{\varphi_1}
{\varphi_0^{\prime} }^{3} {a^\prime }^{2} a^{-4} - 2 {\delta}^{i j}
a^\prime {\partial}_{i}{\varphi_1} {\partial}_{0 j}{B_1}
{\varphi_0^{\prime} }^{3} a^{-3} + \\
&
2 {\delta}^{i j} a^\prime
{\partial}_{i}{\phi_1} {\partial}_{j}{\varphi_1} {\varphi_0^{\prime}
}^{3} a^{-3} + 2 {\delta}^{i j} {\partial}_{i}{B_1} \varphi_0^{\prime
  \prime} {\partial}_{0 j}{B_1} {\varphi_0^{\prime} }^{3} a^{-2} - 2
{\delta}^{i j} {\partial}_{i}{B_1} {\partial}_{j}{\phi_1}
\varphi_0^{\prime \prime} {\varphi_0^{\prime} }^{3} a^{-2} -
\bm{{\phi_1^\prime}^2 {\varphi_0^{\prime} }^{4} a^{-2}} - \\
&
2
{\delta}^{i j} a^\prime {\partial}_{i}{B_1} {\partial}_{0 j}{B_1}
{\varphi_0^{\prime} }^{4} a^{-3} + 2 {\delta}^{i j} a^\prime
{\partial}_{i}{B_1} {\partial}_{j}{\phi_1} {\varphi_0^{\prime} }^{4}
a^{-3} + 2 {\delta}^{i j} {\partial}_{i}{\phi_1}
{\partial}_{j}{\phi_1} {\varphi_0^{\prime} }^{4} a^{-2} - {\delta}^{i
  j} {\delta}^{k l} {\partial}_{i k}{B_1} {\partial}_{j l}{B_1}
{\varphi_0^{\prime} }^{4} a^{-2} - \\
&
{\varphi_0^{\prime} }^{2}
{\varphi_1^{\prime \prime} }^{2} a^{-2} + 2 {\delta}^{i j}
{\partial}_{i 0}{\varphi_1} {\partial}_{j 0}{\varphi_1}
{\varphi_0^{\prime} }^{2} a^{-2} - {\delta}^{i j} {\delta}^{k l}
{\partial}_{k i}{\varphi_1} {\partial}_{l j}{\varphi_1}
{\varphi_0^\prime}^{2} a^{-2} - 4 {\delta}^{i j} a^\prime
{\partial}_{i}{\varphi_1} {\partial}_{j 0}{\varphi_1}
{\varphi_0^{\prime} }^{2} a^{-3} \}
\end{split}
\end{equation}}

It is important to note that the second order action contains
${\phi^\prime}^2$ which cannot be absorbed as a boundary term (the
highlighted term in the above expression). Hence the constraint
condition is not satisfied leading to the fact that the reduced action
does not lead to single variable equation of motion.

From the above analysis, we can generalize and apply this to any
higher derivative scalar theory models like $ \{\Box \varphi\}^3,~
\Box \Box \varphi,~ \varphi^{; \alpha}_{;\beta}
\varphi^{;\beta}_{;\gamma} \varphi^{;\gamma}_{;\alpha}, ~\varphi_{;
  \alpha \beta \gamma \delta} \varphi^{\alpha \beta \gamma \delta}$
and that obtaining a single variable equation of motion or action is
not possible for these kind of models.

\subsection{\texorpdfstring{$f(R)$}{TEXT} model}\label{fr}
Until now, we have considered different forms of scalar field action
without modifying gravity. In this subsection, we consider the
simplest modification i.e. $R + \alpha R^2$ while we consider the
matter to be canonical scalar field. Since $R$ and matter part of the
action does not have any inconsistency, we expand $R^2$ up to second
order in perturbed variables to get,

{\scriptsize
\begin{equation}
\begin{split}
 & 36 {\delta}^{i j} {\delta}^{k l} {\partial}_{i j}{B_1} {\partial}_{k
    l}{B_1} {a^{\prime}}^{2} a^{-6} + 12 {\delta}^{i j} {\delta}^{k l}
  a^{\prime} {\partial}_{i j}{B_1} {\partial}_{0 k l}{B_1} a^{-5} + 72
  {\delta}^{i j} \phi_1^{\prime} {\partial}_{i j}{B_1}
  {a^{\prime}}^{2} a^{-6} + \\
  &
  288 \phi_1 {\delta}^{i j} a^{\prime}
  a^{\prime \prime} {\partial}_{i j}{B_1} a^{-6} + 12 {\delta}^{i j}
  {\delta}^{k l} a^{\prime} {\partial}_{i j}{B_1} {\partial}_{k
    l}{\phi_1} a^{-5} + 12 {\delta}^{i j} {\delta}^{k l} a^{\prime}
  {\partial}_{k l}{B_1} {\partial}_{0 i j}{B_1} a^{-5} + 4 {\delta}^{i
    j} {\delta}^{k l} {\partial}_{0 i j}{B_1} {\partial}_{0 k l}{B_1}
  a^{-4} + \\
  &
  24 {\delta}^{i j} \phi_1^{\prime} a^{\prime} {\partial}_{0
    i j}{B_1} a^{-5} + 96 \phi_1 {\delta}^{i j} a^{\prime \prime}
  {\partial}_{0 i j}{B_1} a^{-5} + 4 {\delta}^{i j} {\delta}^{k l}
  {\partial}_{k l}{\phi_1} {\partial}_{0 i j}{B_1} a^{-4} + \bm{36
    {\phi_1^\prime}^2 {a^{\prime} }^{2} a^{-6}} + \\
    &
    432 \phi_1
  \phi_1^{\prime} a^{\prime} a^{\prime \prime} a^{-6} + 24 {\delta}^{i
    j} \phi_1^{\prime} a^{\prime} {\partial}_{i j}{\phi_1} a^{-5} +
  432 \phi_1^{2} {a^{\prime \prime} }^{2} a^{-6} + 96 \phi_1
  {\delta}^{i j} a^{\prime \prime} {\partial}_{i j}{\phi_1} a^{-5} +\\
  &
  12 {\delta}^{i j} {\delta}^{k l} a^{\prime} {\partial}_{k l}{B_1}
  {\partial}_{i j}{\phi_1} a^{-5} + 4 {\delta}^{i j} {\delta}^{k l}
  {\partial}_{i j}{\phi_1} {\partial}_{0 k l}{B_1} a^{-4} + 4
  {\delta}^{i j} {\delta}^{k l} {\partial}_{i j}{\phi_1} {\partial}_{k
    l}{\phi_1} a^{-4} + 24 {\delta}^{i j} \phi_1^{\prime} a^{\prime
    \prime} {\partial}_{i j}{B_1} a^{-5} - \\
    &
    72 {\delta}^{i j}
  a^{\prime} {\partial}_{i}{B_1} a^{\prime \prime} {\partial}_{0
    j}{B_1} a^{-6} + 12 {\delta}^{i j} {\delta}^{k l} a^{\prime
    \prime} {\partial}_{i j}{B_1} {\partial}_{k l}{B_1} a^{-5} + 72
  {\delta}^{i j} a^{\prime} {\partial}_{i}{B_1} {\partial}_{j}{\phi_1}
  a^{\prime \prime} a^{-6} - 72 {\delta}^{i j} {\partial}_{i}{B_1}
  {\partial}_{j}{B_1} {a^{\prime \prime} }^{2} a^{-6} + \\
  &
  24 {\delta}^{i
    j} {\partial}_{i}{\phi_1} {\partial}_{j}{\phi_1} a^{\prime \prime}
  a^{-5} - 12 {\delta}^{i j} {\delta}^{k l} a^{\prime \prime}
  {\partial}_{i k}{B_1} {\partial}_{j l}{B_1} a^{-5}
  \end{split}
\end{equation}}

Following points are interesting to note from the above expression:
(i) $R^2$ term contains time derivative of $\phi$, that cannot be
absorbed as a boundary term. This implies that the above action cannot
lead to the constraint equation. (ii) By doing a conformal
transformation $g_{\mu \nu} \rightarrow \tilde{g_{\mu \nu}} = \Omega^2
g_{\mu \nu} $, the term containing the time derivative of $\phi$ can
be absorbed as a matter field and hence the constraint equation
recovered. (iii) The constraint consistency allows us to identify that
the $f(R)$ gravity models, without conformal transformation lead to
inconsistent dynamics.

\subsection{\texorpdfstring{$[\varphi_{;\lambda} \varphi^{;\lambda}( \{\Box
  \varphi\}^2 - \varphi_{; \mu \nu} \varphi^{; \mu \nu} )]$}{TEXT} model}
As we have shown above, certain higher derivative models do not
satisfy `constraint consistency'. We have also shown which terms in
the second order action spoil the `constraint consistency'. However,
it is interesting to note the terms that contain time derivative of
Lapse functions are identical. Let us consider the following scalar
field action

\begin{equation}
  S = \int d^4x \sqrt{-g}~[~\varphi_{;\lambda} \varphi^{;\lambda}~\left( \{\Box \varphi\}^2 - \varphi_{; \mu \nu} \varphi^{; \mu \nu} \right)~]
\end{equation}
The second order action is given by,

{\scriptsize
\begin{equation}
\begin{split}
  ^{(2)}\mathcal{S}_m =& \int d^4x \{ - 105 a^{\prime}
  \varphi_0^{\prime \prime} \phi_1^{2} {\varphi_0^{\prime}}^{3} a^{-3}
  - 10 \phi_1 {\delta}^{i j} \varphi_0^{\prime \prime} {\partial}_{i
    j}{B_1} {\varphi_0^{\prime} }^{3} a^{-2} - 42 \phi_1
  \phi_1^{\prime} a^{\prime} {\varphi_0^{\prime} }^{4} a^{-3} - 10
  \phi_1 {\delta}^{i j} a^{\prime} {\partial}_{i j}{B_1}
  {\varphi_0^{\prime} }^{4} a^{-3} - \\
  &
  6 \phi_1 {\delta}^{i j}
  \varphi_0^{\prime \prime} {\partial}_{j i}{\varphi_1}
  {\varphi_0^{\prime} }^{2} a^{-2} + 30 \phi_1 a^{\prime}
  \varphi_1^{\prime \prime} {\varphi_0^{\prime} }^{3} a^{-3} + 90
  \phi_1 {\varphi_1^\prime} a^{\prime} \varphi_0^{\prime \prime}
  {\varphi_0^{\prime} }^{2} a^{-3} - 6 \phi_1 {\delta}^{i j}
  a^{\prime} {\partial}_{j i}{\varphi_1} {\varphi_0^{\prime} }^{3}
  a^{-3} + \\
  &
  6 {\delta}^{i j} {\varphi_1^\prime} \varphi_0^{\prime
    \prime} {\partial}_{i j}{B_1} {\varphi_0^{\prime} }^{2} a^{-2} +
  24 \phi_1^{\prime} {\varphi_1^\prime} a^{\prime} {\varphi_0^{\prime}
  }^{3} a^{-3} + 8 {\delta}^{i j} {\varphi_1^\prime} a^{\prime}
  {\partial}_{i j}{B_1} {\varphi_0^{\prime} }^{3} a^{-3} + 4
  {\delta}^{i j} \varphi_0^{\prime} {\varphi_1^\prime}
  \varphi_0^{\prime \prime} {\partial}_{j i}{\varphi_1} a^{-2} - \\
  &
  18
  {\varphi_1^\prime} a^{\prime} \varphi_1^{\prime \prime}
  {\varphi_0^{\prime} }^{2} a^{-3} - 18 \varphi_0^{\prime} a^{\prime}
  \varphi_0^{\prime \prime} {\varphi_1^{\prime} }^{2} a^{-3} + 6
  {\delta}^{i j} {\varphi_1^\prime} a^{\prime} {\partial}_{j
    i}{\varphi_1} {\varphi_0^{\prime} }^{2} a^{-3} + 15 {\delta}^{i j}
  a^{\prime} {\partial}_{i}{B_1} {\partial}_{j}{B_1} \varphi_0^{\prime
    \prime} {\varphi_0^{\prime} }^{3} a^{-3} + \\
    &
    12 {\delta}^{i j}
  a^{\prime} {\partial}_{j}{B_1} {\partial}_{i}{\varphi_1}
  \varphi_0^{\prime \prime} {\varphi_0^{\prime} }^{2} a^{-3} + 6
  {\delta}^{i j} \varphi_0^{\prime} a^{\prime}
  {\partial}_{i}{\varphi_1} {\partial}_{j}{\varphi_1}
  \varphi_0^{\prime \prime} a^{-3} - 2 {\delta}^{i j}
  {\partial}_{i}{\varphi_1} {\partial}_{j}{\varphi_1}
  {\varphi_0^{\prime} }^{2} {a^{\prime} }^{2} a^{-4}%
  + 6 {\delta}^{i j} a^{\prime} {\partial}_{i}{B_1}
  {\partial}_{j}{\varphi_1} \varphi_0^{\prime \prime}
  {\varphi_0^{\prime} }^{2} a^{-3} + \\
  &
  2 {\delta}^{i j}
  \varphi_1^{\prime \prime} {\partial}_{i j}{B_1} {\varphi_0^{\prime}
  }^{3} a^{-2} + 12 {\delta}^{i j} a^{\prime} {\partial}_{i}{B_1}
  {\partial}_{j 0}{\varphi_1} {\varphi_0^{\prime} }^{3} a^{-3} - 2
  {\delta}^{i j} \phi_1^{\prime} {\partial}_{j i}{\varphi_1}
  {\varphi_0^{\prime} }^{3} a^{-2} - 2 {\delta}^{i j} {\delta}^{k l}
  {\partial}_{i j}{B_1} {\partial}_{l k}{\varphi_1}
  {\varphi_0^{\prime} }^{3} a^{-2} + \\
  &
  6 {\delta}^{i j} a^{\prime}
  {\partial}_{i}{\varphi_1} {\partial}_{0 j}{B_1} {\varphi_0^{\prime}
  }^{3} a^{-3} + 2 {\delta}^{i j} a^{\prime} {\partial}_{i}{\phi_1}
  {\partial}_{j}{\varphi_1} {\varphi_0^{\prime} }^{3} a^{-3} + 6
  {\delta}^{i j} a^{\prime} {\partial}_{i}{B_1} {\partial}_{0 j}{B_1}
  {\varphi_0^{\prime} }^{4} a^{-3} - 6 {\delta}^{i j} a^{\prime}
  {\partial}_{i}{B_1} {\partial}_{j}{\phi_1} {\varphi_0^{\prime} }^{4}
  a^{-3} - \\
  &
  2 {\delta}^{i j} \phi_1^{\prime} {\partial}_{i j}{B_1}
  {\varphi_0^{\prime} }^{4} a^{-2} - {\delta}^{i j} {\delta}^{k l}
  {\partial}_{i j}{B_1} {\partial}_{k l}{B_1} {\varphi_0^{\prime}
  }^{4} a^{{-2}} + 2 {\delta}^{i j} \varphi_1^{\prime \prime}
  {\partial}_{j i}{\varphi_1} {\varphi_0^{\prime} }^{2} a^{{-2}} -
  {\delta}^{i j} {\delta}^{k l} {\partial}_{j i}{\varphi_1}
  {\partial}_{l k}{\varphi_1} {\varphi_0^{\prime} }^{2} a^{-2} + \\
  &
  4
  {\delta}^{i j} {\partial}_{i}{\phi_1} {\partial}_{j 0}{\varphi_1}
  {\varphi_0^{\prime} }^{3} a^{-2} + 2 {\delta}^{i j} {\delta}^{k l}
  {\partial}_{i k}{B_1} {\partial}_{l j}{\varphi_1}
  {\varphi_0^{\prime} }^{3} a^{-2} - 2 {\delta}^{i j}
  {\partial}_{i}{\phi_1} {\partial}_{j}{\phi_1} {\varphi_0^{\prime}
  }^{4} a^{-2} + {\delta}^{i j} {\delta}^{k l} {\partial}_{i k}{B_1}
  {\partial}_{j l}{B_1} {\varphi_0^{\prime} }^{4} a^{-2} - \\
  &
  2
  {\delta}^{i j} {\partial}_{i 0}{\varphi_1} {\partial}_{j
    0}{\varphi_1} {\varphi_0^{\prime} }^{2} a^{-2} + {\delta}^{i j}
  {\delta}^{k l} {\partial}_{k i}{\varphi_1} {\partial}_{l
    j}{\varphi_1} {\varphi_0^{\prime} }^{2} a^{-2} + 4 {\delta}^{i j}
  a^{\prime} {\partial}_{i}{\varphi_1} {\partial}_{j 0}{\varphi_1}
  {\varphi_0^{\prime} }^{2} a^{-3}\}
  \end{split}
\end{equation}}

It is interesting to note the following points: (i) the action does
not contain terms having time derivative of Lapse function/Shift
vector. Hence the resultant equation leads to constraint
equation. Similarly, $[\varphi_{;\lambda} \varphi^{;\lambda}(\{\Box
\varphi\}^3 - 3 \Box \varphi ~\varphi_{; \mu \nu}\varphi^{; \mu \nu} +
2 \varphi^{; \alpha}_{;\beta} \varphi^{;\beta}_{;\gamma}
\varphi^{;\gamma}_{;\alpha})]$ model do not have dynamical
Lapse/Shift. (ii) From the above, one may be tempted to relate the
constraint consistency with Ostrogradsky's instabilities. To go about
checking this, let us look at the zeroth order action for the matter
field, i.e.,

\begin{equation}
  ^{(0)}\mathcal{S}_m = \int d^4x \{- 6 a^{\prime} \varphi_0^{\prime
    \prime} {\varphi_0^{\prime} }^{3} a^{(-3)}\}
\end{equation}
which, after integration by-parts, can be re-written as,

\begin{equation}\label{oszero}
^{(0)}\mathcal{S}_m = \int d^4x \{ \frac{3}{2} a^{\prime \prime}a^{-3} {\varphi_0^{\prime} }^{4} - \frac{9}{2} {a^\prime}^2 a^{-4} {\varphi_0^{\prime} }^{4}\}
\end{equation}
Hence the equations of motion will contain $a^{\prime \prime \prime}$
and $\phi_0^{\prime\prime\prime}$. This immediately signals
Ostrogradsky's instability.

This leads us to the important conclusion: \textit{While all
  `constraint' inconsistent models have higher order Ostrogradsky's
  instabilities but the reverse is not true. One can have models with
  constraint Lapse function and Shift vector, though it may have
  Ostrogradsky's instabilities.}

\subsection{Constraint consistent models without instabilities}
In the previous subsection, we showed that identifying the terms in
(\ref{box}) and (\ref{boxd}) that contains higher derivatives of Lapse
function, we can remove these terms by combination of these terms two
terms. However, we noticed that such actions suffer from
Ostrogradsky's instabilities. The term that leads to Ostrogradsky's
instability is $\frac{3}{2} a^{\prime \prime} a^{-3}
{\varphi_{0}^{\prime}}^4$ at zeroth order action (\ref{oszero}). In
order to cancel such term, without involving any higher derivatives of
$\phi$, one needs to introduce non-minimal coupling of the kinetic
term of the scalar field with Ricci scalar. It is easy to show check
that the term $-\frac{1}{4} (\varphi_{; \alpha} \varphi^{; \alpha})^2
R$ exactly cancels this and the resulting background action becomes,

\begin{equation}
^{(0)}\mathcal{S}_m = - \frac{9}{2} \int d^4x ~ {a^\prime}^2 a^{-4} {\varphi_0^{\prime} }^{4}
\end{equation}

Similarly, $[\varphi_{; \lambda} \varphi^{; \lambda}(\{\Box
\varphi\}^3 - 3 \Box \varphi ~\varphi_{; \mu \nu}\varphi^{; \mu \nu} +
2 \varphi^{; \alpha}_{;\beta} \varphi^{;\beta}_{;\gamma}
\varphi^{;\gamma}_{;\alpha} - 6 G^{\mu \nu} \varphi^{;\alpha}_{;\mu}
\varphi_{; \alpha} \varphi_{;\nu})]$ does not have any constraint
inconsistencies as well as instabilities.

The following points are worth noting regarding this:
\begin{itemize}
\item[i.] We have arrived at the action $[\varphi_{;\lambda}
  \varphi^{;\lambda}(\{\Box \varphi\}^2 - \varphi_{; \mu
    \nu}\varphi^{; \mu \nu} -\frac{1}{4} (\varphi_{; \alpha}
  \varphi^{; \alpha})^2 R]$ by using the condition that the action
  does not have time derivatives of Lapse function and Shift vector
  and later using the additional condition that Ostrogradsky's
  instability does not arise.
\item[ii.] In a different manner, Nicolis et al\citep{Nicolis2008} and
  Deffayet et al\citep{Deffayet2009} have come up with the same
  action, only with condition of removing Ostrogradsky's
  instability. Here we have verified that those models will result in
  consistent dynamics with `constraint consistency' and lead to single
  variable action as well as equation of motion.
\item[iii.]  Similarly, we find that the third order derivative model
  is $[\varphi_{; \lambda} \varphi^{; \lambda}(\{\Box \varphi\}^3 - 3
  \Box \varphi ~\varphi_{; \mu \nu}\varphi^{; \mu \nu} + 2 \varphi^{;
    \alpha}_{;\beta} \varphi^{;\beta}_{;\gamma}
  \varphi^{;\gamma}_{;\alpha} - 6 G^{\mu \nu} \varphi^{;\alpha}_{;\mu}
  \varphi_{; \alpha} \varphi_{;\nu})]$ which, again has been derived
  in \cite{Nicolis2008} and \citep{Deffayet2009}. This model is also
  constraint consistent and free of Ostrogradsky's instability.

\end{itemize}

\section{Conclusions}
In this work, we have revisited the two approaches in cosmological
perturbation theory --- order-by-order approach of the Einstein's
equation and reduced action formalism. Equivalence of the two
approaches were not clear since Gravity equations are highly
non-linear. In this work, we have established that equations arising
from order-by-order approach of the Einstein's equations and those
from the action formalism are equivalent for canonical as well as
non-canonical scalar fields up to second order. These results may
easily be extended to any model in Gravity models at any order.

To compare both the approaches, We have identified a `Constraint
consistency condition' where the constrained nature of Lapse function
and Shift vector are studied. We have shown that, in order to obtain a
reduced equation for both of the approaches, `Constraint consistency'
relation has to be satisfied, i.e., those variables should appear in
the action algebraically, and no non-reducible partial time
derivatives of Lapse function and Shift vector should be present. In
other words, equations of motion of Lapse function and Shift vector
should not contain second order partial time derivatives. We then
investigated the higher order derivative theories of gravity and found
that models which satisfy the constraint conditions can be applied to
the conventional perturbation theory where we express all equations in
a simplified form with a single variable (Curvature perturbation
$\mathcal{R}$ or Mukhanov-Sasaki variable) equation of motion. One
common problem with higher order derivative theories is that, they may
have Higher derivative equations of motion which can give rise to
Ostrogradsky's instabilities. We showed that all the models which do
not satisfy `Constraint consistency' conditions suffer form
Ostrogradsky's instabilities. However, we also find that, there exist
some models which satisfy constraint conditions though they have
instabilities, i.e., the action can be reduced in a single variable
form but the single variable equation of motion will contain higher
order time derivatives. The method we have proposed here is fast and
efficient and would be useful for inflationary model building.

We have also constructed some higher derivative models in such a way
that those models should satisfy constraint consistency condition and
can be free from higher order time derivatives. Those models have
already been derived in the literature in a different manner where the
approach of constructing models are different. This ensures that,
conventional perturbation theory can be used in higher derivative
models which are free from Ostrogradsky's instabilities to obtain a
gauge invariant single variable equation of motion using any
approaches. We summarize the main results,
\begin{enumerate}
\item The two approaches are completely equivalent.
\item Not all models with Lagrangian density $\mathcal{L} = \int
  \sqrt{-g}\left\{R + F(\varphi, \partial\varphi, \partial^2
    \varphi)\right\}$ or $\mathcal{L} = \int \sqrt{-g}
  F(R, \partial\varphi, \partial^2 \varphi)$ can be reduced in a
  single variable form.
\item To obtain a single variable form, `Constraint consistency'
  condition has to be satisfied.
\item Models with constraint inconsistency show Ostrogradsky's
  instability but the reverse is not true.
\end{enumerate}

The analysis here may have implications for models of quantum
cosmology with scalar fields\cite{Bojowald:2008zzb}. The quantum
corrections to the matter and gravity can be modelled as effective
stress-tensor \cite{Sotiriou:2008ya}. The effective classical equation
must also satisfy the constraint consistency. This might help to
constraint the quantum cosmology models\cite{Sotiriou:2008ya}.
\section{Acknowledgements}
We thank Sanil Unnikrishnan for useful discussions. We thank the
support of Max Planck-India Partner group on Gravity and Cosmology.
DN is supported by CSIR fellowship. SS is partially supported by
Ramanujan Fellowship of DST, India. Further we thank Kasper Peeters
for his useful program
Cadabra\cite{Peeters:2007wn,DBLP:journals/corr/abs-cs-0608005} and
very useful algebraic calculations with it.

\appendix

\section{Coefficients of second order equation of motion of
  non-canonical scalar field} \label{App:Appendix A} {\small{
\begin{equation}
\begin{split}
 \qquad C_X = &\varphi_2^{\prime \prime} - 2 \phi_1 \varphi_1^{\prime \prime}
  - \phi_2 \varphi_0^{\prime \prime} - \nabla^2{\varphi_2} - 2
  \phi_1^{\prime} \varphi_1^{\prime} - \phi_2 \varphi_0^{\prime} + 2
  \mathcal{H} \varphi_2^{\prime} + 3\varphi_0^{\prime \prime}
  \phi_1^{2} - 2 \phi_1 \nabla^2{\varphi_1} + 6 \phi_1 \phi_1^{\prime}
  \varphi_0^{\prime} - \\
  &
  4 \mathcal{H} \phi_1 {\varphi_1^{\prime}} - 2
  \mathcal{H} \phi_2 \varphi_0^{\prime} - \varphi_0^{\prime}
  \nabla^2{B_2} - 2 {\varphi_1^{\prime}} \nabla^2{B_1} - 4 {\delta}^{i
    j} {\partial}_{i}{B_1} {\partial}_{j}{\varphi_1^{\prime}} - 2
  {\delta}^{i j} {\partial}_{i}{\phi_1} {\partial}_{j}{\varphi_1} - 2
  {\delta}^{i j} {\partial}_{i}{\varphi_1}
  {\partial}_{j}{B_1^{\prime}} + \\
  &
  6 \mathcal{H} \varphi_0^{\prime}
  \phi_1^{2} + 2 \phi_1 {\varphi_0^{\prime}} \nabla^2{B_1} + 2
  {\delta}^{i j} \varphi_0^{\prime} {\partial}_{i}{B_1}
  {\partial}_{j}{\phi_1} - 2 {\delta}^{i j} {\varphi_0^{\prime}}
  {\partial}_{i}{B_1} {\partial}_{j}{B_1^{\prime}} - 4 \mathcal{H}
  {\delta}^{i j} {\partial}_{i}{B_1} {\partial}_{j}{\varphi_1} -\\
  &
  {\delta}^{i j} {\partial}_{i}{B_1} {\partial}_{j}{B_1}
  {\varphi_0^{\prime \prime}} -
   2 \mathcal{H} {\delta}^{i j}
  {\varphi_0^{\prime}} {\partial}_{i}{B_1} {\partial}_{j}{B_1}
  \end{split}
\end{equation}}}
{\small{
\begin{equation}
\begin{split}
  \qquad C_{XX} =& a^{-2}\phi_2^{\prime} {\varphi_0^{\prime}}^{3} - 3 a^{-2}
  {\varphi_0^{\prime \prime}} {\varphi_1^{\prime}}^{2} - a^{-2}
  {\varphi_2^{\prime \prime}}{\varphi_0^{\prime}}^{2} - 12 a^{-2}
  \phi_1 \phi_1^{\prime} {\varphi_0^{\prime}}^{3} + 8 a^{-2} \phi_1
  {\varphi_1^{\prime \prime}} {\varphi_0^{\prime}}^{2} + \\
  &
  4 a^{-2}
  \phi_2 {\varphi_0^{\prime \prime}} {\varphi_0^{\prime}}^{2} + 8
  a^{-2} {\phi_1^{\prime}}
  {\varphi_1^{\prime}}{\varphi_0^{\prime}}^{2} - 6 a^{-2}
  {\varphi_0^{\prime}}{\varphi_1^{\prime}}{\varphi_1^{\prime \prime}}
  - 3 a^{-2} {\varphi_0^{\prime}}
  {\varphi_2^{\prime}}{\varphi_0^{\prime \prime}} - 21 a^{-2}
  {\varphi_0^{\prime \prime}} \phi_1^{2} {\varphi_0^{\prime}}^{2} - \\
  &
  2
  a^{-2} \phi_1 \nabla^2{B_1}{\varphi_0^{\prime}}^{3} - 2 a^{-2}
  \phi_1 \nabla^2{\varphi_1} {\varphi_0^{\prime}}^{2} + 18 a^{-2}
  \phi_1 {\varphi_0^{\prime}} {\varphi_1^{\prime}} {\varphi_0^{\prime
      \prime}} - a^{-2} \phi_2 \mathcal{H} {\varphi_0^{\prime}}^{3} +
  2 a^{-2} {\varphi_0^{\prime}}{\varphi_1^{\prime}}
  \nabla^2{\varphi_1} + \\
  &
  4 a^{-2} {\delta}^{i j} {\varphi_0}
  {\partial}_{i}{\varphi_1} {\partial}_{j}{\varphi_1^{\prime}} + 2
  a^{-2} {\varphi_1^{\prime}} \nabla^2{B_1} {\varphi_0^{\prime}}^{2} -
  2 a^{-2} {\delta}^{i j} {\partial}_{i}{B_1} {\partial}_{j}{\phi_1}
  {\varphi_0^{\prime}}^{3} + \\
  &
  2 a^{-2} {\delta}^{i j}
  {\partial}_{i}{B_1}
  {\partial}_{j}{B_1^{\prime}}{\varphi_0^{\prime}}^{3} + 4 a^{-2}
  {\delta}^{i j} {\partial}_{i}{B_1}
  {\partial}_{j}{\varphi_1^{\prime}} {\varphi_0^{\prime}}^{2} -
   2
  a^{-2} {\delta}^{i j} {\partial}_{i}{\phi_1}
  {\partial}_{j}{\varphi_1} {\varphi_0^{\prime}}^{2} + a^{-2}
  {\delta}^{i j} {\partial}_{i}{\varphi_1} {\partial}_{j}{\varphi_1}
  {\varphi_0^{\prime \prime}} + \\
  &
  2 a^{-2} {\delta}^{i j}
  {\partial}_{i}{\varphi_1} {\partial}_{j}{B_1^{\prime}}
  {\varphi_0^{\prime}}^{2} + 
  a^{-2} \mathcal{H} \varphi_2^{\prime}
  {\varphi_0^{\prime}}^{2} + 3 a^{-2} \mathcal{H} \phi_1^{2}
  {\varphi_0^{\prime}}^{3} - 
  2 a^{-2} \mathcal{H} \phi_1
  {\varphi_1^{\prime}} {\varphi_0^{\prime}}^{2} + \\
  &
  6 a^{-2} {\delta}^{i
    j} {\varphi_0^{\prime}} {\partial}_{i}{B_1}
  {\partial}_{j}{\varphi_1} {\varphi_0^{\prime \prime}} + 4 a^{-2}
  {\delta}^{i j} {\partial}_{i}{B_1} {\partial}_{j}{B_1}
  {\varphi_0^{\prime \prime}} {\varphi_0^{\prime}}^{2} -
   a^{-2}
  \mathcal{H} {\delta}^{i j} {\partial}_{i}{B_1} {\partial}_{j}{B_1}
  {\varphi_0^{\prime}}^{3} -\\
  &
   2 a^{-2} \mathcal{H} {\delta}^{i j}
  {\partial}_{i}{B_1} {\partial}_{j}{\varphi_1}
  {\varphi_0^{\prime}}^{2}
  \end{split}
\end{equation}}}

{\small{
\begin{equation}
\begin{split}
 \qquad C_{XXX} =& 2 a^{-4} \phi_1 {\phi_1^{\prime}} {\varphi_0^{\prime}}^{5}
  - 2 a^{-4} \phi_1 {\varphi_1^{\prime
      \prime}}{\varphi_0^{\prime}}^{4} + a^{-4} \mathcal{H} \phi_2
  {\varphi_0^{\prime}} ^{5} - a^{-4} \phi_2 \varphi_0^{\prime \prime}
  {\varphi_0^{\prime}}^{4} - 2 a^{-4} \phi_1^{\prime}
  \varphi_1^{\prime} {\varphi_0^{\prime}} ^{4} + \\
  &
  2 a^{-4}
  \varphi_1^{\prime} \varphi_1^{\prime \prime}
  {\varphi_0^{\prime}}^{3} - a^{-4} \mathcal{H}\varphi_2^{\prime}
  {\varphi_0^{\prime}}^{4} + a^{-4} \varphi_2^{\prime}
  \varphi_0^{\prime \prime} {\varphi_0^{\prime}}^{3} - 8 a^{-4}
  \mathcal{H} \phi_1^{2} {\varphi_0^{\prime}} ^{5} - 5 a^{-4}
  \mathcal{H} {\varphi_0^{\prime}}^{3} {\varphi_1^{\prime}}^{2} + \\
  &
  11
  a^{-4} \varphi_0^{\prime \prime} \phi_1^{2} {\varphi_0^{\prime}}
  ^{4} + 6 a^{-4} \varphi_0^{\prime \prime} {\varphi_0^{\prime}}^{2}
  {\varphi_1^{\prime}} ^{2} + 12 a^{-4} \mathcal{H} \phi_1
  \varphi_1^{\prime} {\varphi_0^{\prime}}^{4} - 16 a^{-4} \phi_1
  \varphi_1^{\prime} \varphi_0^{\prime \prime}
  {\varphi_0^{\prime}}^{3} + \\
  &
  a^{-4} \mathcal{H} {\delta}^{i j}
  {\partial}_{i}{B_1} {\partial}_{j}{B_1} {\varphi_0^{\prime}}^{5} +
  2
  a^{-4} \mathcal{H} {\delta}^{i j} {\partial}_{i}{B_1}
  {\partial}_{j}{\varphi_1} {\varphi_0^{\prime}} ^{4} + a^{-4}
  \mathcal{H} {\delta}^{i j} {\partial}_{i}{\varphi_1}
  {\partial}_{j}{\varphi_1} {\varphi_0^{\prime}}^{3} - \\
  &
  a^{-4}
  {\delta}^{i j} {\partial}_{i}{B_1} {\partial}_{j}{B_1}
  \varphi_0^{\prime \prime} {\varphi_0^{\prime}} ^{4}
  - 2 a^{-4}
  {\delta}^{i j} {\partial}_{i}{B_1} {\partial}_{j}{\varphi_1}
  \varphi_0^{\prime \prime} {\varphi_0^{\prime}}^{3} -
  a^{-4}
  {\delta}^{i j} {\partial}_{i}{\varphi_1} {\partial}_{j}{\varphi_1}
  \varphi_0^{\prime \prime} {\varphi_0^{\prime} }^{2}
  \end{split}
\end{equation}}}
{\small{
\begin{equation}
\begin{split}
 C_{XXXX} =& a^{-6} \mathcal{H} \phi_1^{2} {\varphi_0^{\prime}}^{7} +
  a^{-6} \mathcal{H} {\varphi_0^{\prime}}^{5} {\varphi_1^{\prime}}^{2}
  - a^{-6} {\varphi_0^{\prime \prime}} \phi_1^{2}
  {\varphi_0^{\prime}}^{6} - a^{-6} {\varphi_0^{\prime \prime}}
  {\varphi_0^{\prime}}^{4} {\varphi_1^{\prime}}^{2} - \\
  &
  2
  a^{-6}\mathcal{H} \phi_1 \varphi_1^{\prime} {\varphi_0^{\prime}}^{6}
  + 2 a^{-6} \phi_1 \varphi_1^{\prime} {\varphi_0^{\prime \prime}}
  {\varphi_0^{\prime}}^{5}
  \end{split}
\end{equation}}}
{\small{
\begin{equation}
\begin{split}
  C_{XXX \varphi} =& a^{-4} \phi_1^{2} {\varphi_0^{\prime}}^{6} +
  a^{-4} {\varphi_0^{\prime}} ^{4} {\varphi_1^{\prime}}^{2} - 2 a^{-4}
  \phi_1 {\varphi_1^{\prime}} {\varphi_0^{\prime}} ^{5} + 2 a^{-4}
  \mathcal{H} \phi_1 {\varphi_0^{\prime}}^{5} \varphi_1 - 2 a^{-4}
  \phi_1 \varphi_0^{\prime \prime} {\varphi_0^{\prime}}^{4} \varphi_1
  - \\
  &
  2 a^{-4} \mathcal{H} {\varphi_1^{\prime}} {\varphi_0^{\prime}}
  ^{4} \varphi_1 + 2 a^{-4} {\varphi_1^{\prime}} \varphi_0^{\prime
    \prime} {\varphi_0^{\prime}}^{3} \varphi_1
    \end{split}
\end{equation}}}
{\small{
\begin{equation}
\begin{split}
  C_{XX\varphi} =& a^{-2} \phi_2 {\varphi_0^{\prime}}^{4} - a^{-2}
  \varphi_2^{\prime \prime} {\varphi_0^{\prime}} ^{3} - 5 a^{-2}
  \phi_1^{2} {\varphi_0^{\prime}} ^{4} - 4 a^{-2} {\varphi_0^{\prime}}
  ^{2} {\varphi_1^{\prime}}^{2} + 8 a^{-2} \phi_1 \varphi_1^{\prime}
  {\varphi_0^{\prime}} ^{3} + \\
  &
  2 a^{-2} {\phi_1^{\prime}}
  {\varphi_0^{\prime}}^{3} \varphi_1 - a^{-2} \varphi_0^{\prime
    \prime} {\varphi_0^{\prime}} ^{2} \varphi_2 - 2 a^{-2}
  \varphi_1^{\prime \prime} {\varphi_0^{\prime}} ^{2} \varphi_1 + 8
  a^{-2} \phi_1 \varphi_0^{\prime \prime} {\varphi_0^{\prime}} ^{2}
  \varphi_1 + \\
  &
  a^{-2} {\delta}^{i j} {\partial}_{i}{B_1}
  {\partial}_{j}{B_1} {\varphi_0^{\prime}} ^{4} + 2 a^{-2} {\delta}^{i
    j} {\partial}_{i}{B_1} {\partial}_{j}{\varphi_1}
  {\varphi_0^{\prime}} ^{3} + a^{-2} {\delta}^{i j}
  {\partial}_{i}{\varphi_1} {\partial}_{j}{\varphi_1}
  {\varphi_0^{\prime}} ^{2} - 6 a^{-2} \varphi_0^{\prime}
  \varphi_1^{\prime} \varphi_0^{\prime \prime} \varphi_1 +\\
  &
   a^{-2}
  \mathcal{H} {\varphi_0^{\prime}} ^{3} \varphi_2 - 2 a^{-2}
  \mathcal{H} \phi_1 {\varphi_0^{\prime}} ^{3} \varphi_1 + 2 a^{-2}
  \mathcal{H} \varphi_1^{\prime} {\varphi_0^{\prime}} ^{2} \varphi_1
  \end{split}
\end{equation}}}

{\small{
\begin{equation}
  C_{XX \varphi \varphi}=2 a^{-2} \phi_1 {\varphi_0^{\prime}}^{4}
  \varphi_1 - 2 a^{-2} {\varphi_1^{\prime}} {\varphi_0^{\prime}}^{3}
  \varphi_1 - a^{-2} {\varphi_0^{\prime \prime}}
  {\varphi_0^{\prime}}^{2} \varphi_1^{2} + a^{-2} \mathcal{H}
  {\varphi_0^{\prime}}^{3} \varphi_1^{2}
\end{equation}
}}

{\small{
\begin{equation}
\begin{split}
  C_{X\varphi} =& {\varphi_1^{\prime}} ^{2} + \varphi_0^{\prime}
  \varphi_2^{\prime} + \varphi_0^{\prime \prime} \varphi_2 + 2
  \varphi_1^{\prime \prime} \varphi_1 + \phi_1^{2}
  {\varphi_0^{\prime}}^{2} - 2 \phi_1 \varphi_0^{\prime}
  \varphi_1^{\prime} -
  2 \phi_1 \varphi_0^{\prime \prime} \varphi_1 -\\
  &
  {\delta}^{i j} {\partial}_{i}{\varphi_1} {\partial}_{j}{\varphi_1} -
  2 {\delta}^{i j} {\partial}_{i j}{\varphi_1} \varphi_1 - 2
  {\phi_1^{\prime}} \varphi_0^{\prime} \varphi_1 + 
  2 \mathcal{H}
  \varphi_0^{\prime} \varphi_2 + 4 \mathcal{H} \varphi_1^{\prime}
  \varphi_1 -\\
  &
   4 \mathcal{H} \phi_1 \varphi_0^{\prime} \varphi_1 - 2
  {\delta}^{i j} \varphi_0^{\prime} {\partial}_{i}{B_1}
  {\partial}_{j}{\varphi_1} - 2 {\delta}^{i j} \varphi_0^{\prime}
  {\partial}_{i j}{B_1} \varphi_1
  \end{split}
\end{equation}
}}

{\small{
\begin{equation}
  C_{X\varphi\varphi} = {\varphi_0^{\prime \prime}} \varphi_1^{2} +
  {\varphi_0^{\prime}}^{2} \varphi_2 + 2 \varphi_0^{\prime}
  {\varphi_1^{\prime}} \varphi_1 + 2 \mathcal{H} {\varphi_0^{\prime}}
  \varphi_1^{2}
\end{equation}
}}

{\small{
\begin{equation}
  C_{X\varphi\varphi\varphi}={\varphi_0^{\prime}}^{2} \varphi_1^{2}
\end{equation}
}}

{\small{
\begin{equation}
  C_{\varphi} = a^{2} \phi_2 - a^{2} \phi_1^{2} + a^{2} {\delta}^{i j}
  {\partial}_{i}{B_1} {\partial}_{j}{B_1}
\end{equation}
}}

\begin{equation}
  C_{\varphi \varphi} = a^{2} \varphi_2 + 2 a^{2} \phi_1 \varphi_1
\end{equation}

\begin{equation}
  C_{\varphi\varphi\varphi} = a^{2} \varphi_1^{2}
\end{equation}

\section{Second order single variable equation of motion for
  non-canonical scalar field for Power-law
  inflation} \label{App:Appendix B} Fourth order action for
non-canonical scalar field, after partial integration is,

{\scriptsize
\begin{equation}\label{nonc}
\begin{split}
  ^{(4)}\mathcal{S} =& \int d^4x \Big( 3 \phi_1 \phi_2 {\delta}^{i j}
  {a^\prime} {\partial}_{i j}{B_1} \kappa^{-1} a + \frac{1}{8} P_{X}
  \phi_2^{2} {\varphi_0^\prime}^{2} a^{2} + \frac{1}{4} P \phi_2^{2}
  a^{4} + \frac{1}{4} P_{X} \phi_2 {\delta}^{i j} {\partial}_{i}{B_1}
  {\partial}_{j}{B_1} {\varphi_0^\prime}^{2} a^{2} + 
  \frac{1}{2} P
  \phi_2 {\delta}^{i j} {\partial}_{i}{B_1} {\partial}_{j}{B_1} a^{4}
  - \\
  &
  \frac{3}{2} P_{X} \phi_2 \phi_1^{2} {\varphi_0^\prime}^{2} a^{2} -
  3 P \phi_2 \phi_1^{2} a^{4} - \frac{1}{4} \phi_2 {\delta}^{i j}
  {\delta}^{k l} {\partial}_{i k}{B_1} {\partial}_{j l}{B_1}
  \kappa^{-1} a^{2} + \frac{1}{4} \phi_2 {\delta}^{i j} {\delta}^{k l}
  {\partial}_{i j}{B_1} {\partial}_{k l}{B_1} \kappa^{-1} a^{2} -
  \frac{3}{2} P_{X} \phi_1 \phi_2 {\varphi_0^\prime}
  {\varphi_1^\prime} a^{2} + \\
  &
  \frac{1}{4} P_{X} \phi_2
  {\varphi_0^\prime} {\varphi_2^\prime} a^{2} + \frac{1}{4} P_{X}
  {\delta}^{i j} {\varphi_0^\prime} {\varphi_2^\prime}
  {\partial}_{i}{B_1} {\partial}_{j}{B_1} a^{2} - \frac{3}{4} P_{X}
  {\varphi_0^\prime} {\varphi_2^\prime} \phi_1^{2} a^{2} + \frac{1}{4}
  P_{X} \phi_2 {\varphi_1^\prime}^{2} a^{2} + \frac{1}{2} P_{X} \phi_1
  {\varphi_1^\prime} {\varphi_2^\prime} a^{2} - \frac{1}{8} P_{X}
  {\varphi_2^\prime}^{2} a^{2} - \\
  &
  \frac{1}{2} P_{X} \phi_2 {\delta}^{i
    j} {\varphi_0^\prime} {\partial}_{i}{B_1}
  {\partial}_{j}{\varphi_1} a^{2} - \frac{1}{2} P_{X} \phi_1
  {\delta}^{i j} {\varphi_0^\prime} {\partial}_{i}{B_1}
  {\partial}_{j}{\varphi_2} a^{2} + \frac{1}{2} P_{X} {\delta}^{i j}
  {\varphi_1^\prime} {\partial}_{i}{B_1} {\partial}_{j}{\varphi_2}
  a^{2}
  + \frac{1}{2} P_{X} {\delta}^{i j} {\varphi_2^\prime}
  {\partial}_{i}{B_1} {\partial}_{j}{\varphi_1} a^{2} + \\
  &
  \frac{1}{8}
  P_{X} {\delta}^{i j} {\partial}_{i}{\varphi_2}
  {\partial}_{j}{\varphi_2} a^{2} - \frac{9}{4} P_{XX} \phi_2
  \phi_1^{2} {\varphi_0^\prime}^{4} + 3 P_{XX} \phi_1 \phi_2
  {\varphi_1^\prime} {\varphi_0^\prime}^{3} + \frac{3}{2} P_{XX}
  {\varphi_2^\prime} \phi_1^{2} {\varphi_0^\prime}^{3} -
  2 P_{XX}
  \phi_1 {\varphi_1^\prime} {\varphi_2^\prime} {\varphi_0^\prime}^{2}
  + \\
  &
  \frac{1}{2} P_{XX} \phi_1 {\delta}^{i j} {\partial}_{i}{B_1}
  {\partial}_{j}{\varphi_2} {\varphi_0^\prime} ^{3} + \frac{1}{2}
  P_{XX} \phi_1 {\delta}^{i j} {\partial}_{i}{\varphi_1}
  {\partial}_{j}{\varphi_2} {\varphi_0^\prime}^{2} + \frac{1}{8}
  P_{XX} \phi_2^{2} {\varphi_0^\prime}^{4} + \frac{1}{4} P_{XX} \phi_2
  {\delta}^{i j} {\partial}_{i}{B_1} {\partial}_{j}{B_1}
  {\varphi_0^\prime}^{4} - \frac{1}{4} P_{XX} \phi_2
  {\varphi_2^\prime} {\varphi_0^\prime}^{3} - \\
  &
  P_{XX} \phi_2 {\varphi_0^\prime}^{2} {\varphi_1^\prime}^{2} + \frac{1}{2} P_{XX}
  \phi_2 {\delta}^{i j} {\partial}_{i}{B_1} {\partial}_{j}{\varphi_1}
  {\varphi_0^\prime}^{3} + \frac{1}{4} P_{XX} \phi_2 {\delta}^{i j}
  {\partial}_{i}{\varphi_1} {\partial}_{j}{\varphi_1}
  {\varphi_0^\prime}^{2} - \frac{1}{4} P_{XX} {\delta}^{i j}
  {\varphi_2^\prime} {\partial}_{i}{B_1} {\partial}_{j}{B_1}
  {\varphi_0^\prime} ^{3} + \\
  &
  \frac{3}{4} P_{XX} {\varphi_0^\prime}
  {\varphi_2^\prime} {\varphi_1^\prime}^{2} - \frac{1}{2} P_{XX}
  {\delta}^{i j} {\varphi_1^\prime} {\partial}_{i}{B_1}
  {\partial}_{j}{\varphi_2} {\varphi_0^\prime}^{2} - \frac{1}{2}
  P_{XX} {\delta}^{i j} {\varphi_0^\prime} {\varphi_1^\prime}
  {\partial}_{i}{\varphi_1} {\partial}_{j}{\varphi_2} + \frac{1}{8}
  P_{XX} {\varphi_0^\prime} ^{2} {\varphi_2^\prime}^{2} - \frac{1}{2}
  P_{XX} {\delta}^{i j} {\varphi_2^\prime} {\partial}_{i}{B_1}
  {\partial}_{j}{\varphi_1} {\varphi_0^\prime}^{2}\\
  &
  - \frac{1}{4} P_{XX} {\delta}^{i j} {\varphi_0^\prime}
  {\varphi_2^\prime} {\partial}_{i}{\varphi_1}
  {\partial}_{j}{\varphi_1} + \frac{1}{8} P_{YY} \varphi_2^{2} a^{4} +
  \frac{1}{4} P_{XY} \phi_2 {\varphi_0^\prime} ^{2} a^{2} \varphi_2 +
  \frac{1}{4} P_{XY} {\delta}^{i j} {\partial}_{i}{B_1}
  {\partial}_{j}{B_1} {\varphi_0^\prime} ^{2} a^{2} \varphi_2 -
  \frac{1}{2} P_{XY} \phi_1^{2} {\varphi_0^\prime}^{2} a^{2} \varphi_2  \\
  &
  - P_{XY} \phi_1 \phi_2 {\varphi_0^\prime} ^{2} a^{2} \varphi_1 +
  \frac{1}{2} P_{XY} \phi_1 {\varphi_0^\prime} {\varphi_1^\prime}
  a^{2} \varphi_2 + \frac{1}{2} P_{XY} \phi_2 {\varphi_0^\prime}
  {\varphi_1^\prime} a^{2} \varphi_1 + \frac{1}{2} P_{XY} \phi_1
  {\varphi_0^\prime} {\varphi_2^\prime} a^{2} \varphi_1 - \frac{1}{4}
  P_{XY} {\varphi_0^\prime} {\varphi_2^\prime} a^{2} \varphi_2 -\\
  &
  \frac{1}{4} P_{XY} {\varphi_1^\prime}^{2} a^{2} \varphi_2 -
  \frac{1}{2} P_{XY} {\varphi_1^\prime} {\varphi_2^\prime} a^{2}
  \varphi_1 + \frac{1}{2} P_{XY} {\delta}^{i j} {\varphi_0^\prime}
  {\partial}_{i}{B_1} {\partial}_{j}{\varphi_1} a^{2} \varphi_2 +
  \frac{1}{2} P_{XY} {\delta}^{i j} {\varphi_0^\prime}
  {\partial}_{i}{B_1} {\partial}_{j}{\varphi_2} a^{2} \varphi_1 +
  \frac{1}{4} P_{XY} {\delta}^{i j} {\partial}_{i}{\varphi_1}
  {\partial}_{j}{\varphi_1} a^{2} \varphi_2 + \\
  &
  \frac{1}{2} P_{XY}
  {\delta}^{i j} {\partial}_{i}{\varphi_1} {\partial}_{j}{\varphi_2}
  a^{2} \varphi_1 + \frac{1}{4} P_{XXX} \phi_2 \phi_1^{2}
  {\varphi_0^\prime}^{6} a^{(-2)} - \frac{1}{4} P_{XXX}
  {\varphi_2^\prime} \phi_1^{2} {\varphi_0^\prime} ^{5} a^{(-2)} -
  \frac{1}{2} P_{XXX} \phi_1 \phi_2 {\varphi_1^\prime}
  {\varphi_0^\prime}^{5} a^{(-2)} + \\
  &
  \frac{1}{2} P_{XXX} \phi_1
  {\varphi_1^\prime} {\varphi_2^\prime} {\varphi_0^\prime} ^{4}
  a^{(-2)}
  + \frac{1}{4} P_{XXX} \phi_2 {\varphi_0^\prime}^{4}
  {\varphi_1^\prime}^{2} a^{(-2)} - \frac{1}{4} P_{XXX}
  {\varphi_2^\prime} {\varphi_0^\prime}^{3} {\varphi_1^\prime}^{2}
  a^{(-2)} + \frac{1}{4} P_{YYY} \varphi_1^{2} a^{4} \varphi_2 +
  \frac{1}{4} P_{XXY} \phi_1^{2} {\varphi_0^\prime}^{4} \varphi_2 +\\
  &
  \frac{1}{2} P_{XXY} \phi_1 \phi_2 {\varphi_0^\prime}^{4} \varphi_1 -
  \frac{1}{2} P_{XXY} \phi_1 {\varphi_1^\prime} {\varphi_0^\prime}
  ^{3} \varphi_2 - \frac{1}{2} P_{XXY} \phi_1 {\varphi_2^\prime}
  {\varphi_0^\prime}^{3} \varphi_1 - \frac{1}{2} P_{XXY} \phi_2
  {\varphi_1^\prime} {\varphi_0^\prime}^{3} \varphi_1 + \frac{1}{4}
  P_{XXY} {\varphi_0^\prime} ^{2} {\varphi_1^\prime}^{2} \varphi_2 +\\
  &
  \frac{1}{2} P_{XXY} {\varphi_1^\prime} {\varphi_2^\prime}
  {\varphi_0^\prime}^{2} \varphi_1 + \frac{1}{2} P_{XYY} \phi_1
  {\varphi_0^\prime}^{2} a^{2} \varphi_1 \varphi_2 + \frac{1}{4}
  P_{XYY} \phi_2 {\varphi_0^\prime}^{2} \varphi_1^{2} a^{2} -
  \frac{1}{2} P_{XYY} {\varphi_0^\prime} {\varphi_1^\prime} a^{2}
  \varphi_1 \varphi_2 - \frac{1}{4} P_{XYY} {\varphi_0^\prime}
  {\varphi_2^\prime} \varphi_1^{2} a^{2} + \\
  &
  \frac{1}{2} P_{X} \phi_1
  {\delta}^{i j} {\partial}_{i}{\varphi_1} {\partial}_{j}{\varphi_2}
  a^{2} + \frac{1}{2} P_{YY} \phi_1 a^{4} \varphi_1 \varphi_2 +
  \frac{1}{4} P_{X} \phi_2 {\delta}^{i j} {\partial}_{i}{\varphi_1}
  {\partial}_{j}{\varphi_1} a^{2} + \frac{1}{4} P_{Y} \phi_2 a^{4}
  \varphi_2 + \frac{1}{4} P_{YY} \phi_2 \varphi_1^{2} a^{4} +\\
  &
  \frac{1}{4} P_{Y} {\delta}^{i j} {\partial}_{i}{B_1}
  {\partial}_{j}{B_1} a^{4} \varphi_2%
  - \frac{1}{4} P_{Y} \phi_1^{2} a^{4} \varphi_2 - \frac{1}{2} P_{Y}
  \phi_1 \phi_2 a^{4} \varphi_1 \Big)
  \end{split}
\end{equation}
}

\noindent We can eliminate $\phi_2,~ \phi_1$ and $B_1$ using
constraint equations. While this is possible for canonical scalar
field, it is highly non-trivial for non-canonical scalar fields. Hence
we consider Power law inflation to reduce the action in a single
variable form.

For Power-law, we have

\begin{eqnarray}
&&a = a_0 (-\eta)^\beta \\
&&\mathcal{H} \equiv \frac{a^\prime}{a}= - \frac{\beta}{(-\eta)}\\
&&\frac{a^{\prime \prime}}{a} = - \frac{\beta (1 -\beta)}{(-\eta)^2} \\
&&\varphi_0 = \varphi_{c} (-\eta)^\alpha \\
&&{\varphi_0}^\prime = \varphi_{c} \alpha (-\eta)^\alpha
\end{eqnarray}

\begin{eqnarray}
  &&\mathcal{H}^2 = -\frac{\kappa}{3} (P_X {{\varphi_0}^{\prime 2}} + P_0 a^2) \\
  &&- 2 \frac{a^{\prime \prime}}{a} + \mathcal{H}^2 = \kappa P_0 a^2
\end{eqnarray}

For Power-law inflation, $P(X, \varphi) = \frac{C}{\varphi^2} ~(X^2 +
X)$. Solving for $C, \varphi_{c}$ and $\alpha$ using above equations,
\begin{eqnarray}
  &&\alpha = 1 + \beta \\
  &&C = \frac{- 6 \beta (1 - \beta)}{\kappa (1 + \beta)^2} \\
  &&\varphi_{0c} = \sqrt{\frac{2}{3}} \frac{a_0}{1 + \beta} \sqrt{\frac{1- 2 \beta}{1- \beta}}
\end{eqnarray}
Using first order Einstein's equations,
\begin{eqnarray}
  &&\phi_1 = \sqrt{\frac{3}{2}} \sqrt{\frac{1- \beta}{1-2\beta}} \frac{1 + \beta}{a_0(-\eta)^{1 + \beta}} \varphi_1 \\
  &&\phi_1 = F_3 \varphi_1 \\
  &&\nabla^2 B_1 = \frac{3}{a_0} \sqrt{\frac{3}{2}} \sqrt{\frac{1- \beta}{1-2\beta}} (1-3\beta) \Big\{ \frac{1 + \beta}{(-\eta)^{2 + \beta}} \varphi_1 + \frac{1}{(-\eta)^{ 1 + \beta}} \varphi_1^\prime \Big\}\\
  &&\nabla^2 B_1 = F_1 \varphi_1 + F_2 \varphi_1^\prime
\end{eqnarray}
where,
\begin{eqnarray}
  &&F_1 = \frac{3}{a_0} \sqrt{\frac{3}{2}} \sqrt{\frac{1- \beta}{1-2\beta}} 
(1-3\beta)\frac{1 + \beta}{(-\eta)^{2 + \beta}} \\
  &&F_2 = \frac{3}{a_0} \sqrt{\frac{3}{2}} \sqrt{\frac{1- \beta}{1-2\beta}} 
(1-3\beta)\frac{1}{(-\eta)^{ 1 + \beta}} \\
  &&F_3 = \frac{1}{a_0}\sqrt{\frac{3}{2}} \sqrt{\frac{1- \beta}{1-2\beta}} 
\frac{1 + \beta}{(-\eta)^{1 + \beta}}
\end{eqnarray}

Similarly using second order Einstein's equations, 
\begin{eqnarray}
  &&\partial_{i}{\phi_2} = C_1 \partial_{i}{\varphi_2} + C_2 \varphi_1 
\partial_{i}{\varphi_1} + C_3 \delta^{j k} \partial_{j}{\varphi_1} 
\partial_{i k}{B_1} + C_4 \delta^{j k} \partial_{j}{B_1} \partial_{i k}{B_1}+ 
C_5 \varphi_1^\prime \partial_{i}{\varphi_1} ~~~~~~~~\\
  &&\phi_2 = C_1 \varphi_2 + \frac{1}{2} C_2 \varphi_1^2 - \delta^{i j} 
\partial_{i}{B_1} \partial_{j}{B_1} + Q \\
  &&Q = {\partial^{-1 i}} \{C_3 \delta^{j k} \partial_{j}{\varphi_1} 
\partial_{i k}{B_1} + C_5 \varphi_1^\prime \partial_{i}{\varphi_1}\} 
\end{eqnarray}
where
\begin{eqnarray}
  &&C_1 = \sqrt{\frac{3}{2}} \sqrt{\frac{1- \beta}{1-2\beta}} 
\frac{1 + \beta}{a_0~(-\eta)^{1 + \beta}}\\
  &&C_2 = - \frac{3}{2 a_0^2} \frac{(1+\beta) 
(1-\beta)(3 - 14\beta +7\beta^2)}{\ (1-2\beta)}\frac{1}{(-\eta)^{2+ 2\beta}}\\
  &&C_3 = \sqrt{\frac{3}{2}}\sqrt{\frac{1-\beta}{1-2\beta}}
\frac{(1+ \beta)}{a_0 \beta} \frac{1}{(-\eta)^{\beta}}\\
  &&C_4 = -2\\
  &&C_5 = - \frac{9}{2 a_0^2} \frac{(1-\beta)^2 (1-3\beta)}{\beta(1-2\beta)}
\frac{1}{(-\eta)^{1+2\beta}}
\end{eqnarray}
Multiplying the fourth order action (\ref{nonc}) by $\frac{8}{3} (1- 2\beta)^2$,
\begin{equation}
\begin{split}
  &\int d^4x \Big[ \Big(A_1 \varphi_2^{\prime 2} +
  A_2 \partial^{i}{\varphi_2} \partial_{i}{\varphi_2} + A_3 \varphi_2
  \varphi_2^\prime + A_4 \varphi_2^2 \Big) + \partial^{i}{\varphi_2}
  ~\Big(B_1 \varphi_1 \partial_{i}{\varphi_1} +
   B_2
  \varphi_1 \partial_{i}{B_1}+ \\
  &
  B_3\varphi_1^\prime \partial_{i}{B_1} +
  B_4 \varphi_1^\prime \partial_{i}{\varphi_1} \Big) +
  \varphi_2^\prime~ \Big(D_1 \varphi_1^2 + D_2 \varphi_1^\prime
  \varphi_1 + D_3 Q + D_4 {\varphi_1^\prime}^2 +
  D_5 \partial^{i}{B_1} \partial_{i}{\varphi_1} +\\
  &
  D_6 \partial^{i}{\varphi_1}\partial_{i}{\varphi_1}\Big) + \varphi_2
  ~ \Big(E_1 \varphi_1^2 + E_2 (\nabla^2 B_1)^2 + E_3 \partial^{i
    j}{B_1} \partial_{i j}{B_1} + E_4 \varphi_1 \varphi_1^\prime + E_5
  {\varphi_1^\prime}^2 + E_6 Q \\
  &
  +
  E_7 \partial^{i}{B_1} \partial_{i}{\varphi_1} +
  E_8 \partial^{i}{\varphi_1} \partial_{i}{\varphi_1} \Big) \Big]
  \end{split}
\end{equation}
where,
\begin{eqnarray}
  &&A_1 = -  \frac{3\beta (1 -\beta)(1-2\beta)(1-3\beta)}{\kappa (-\eta)^2} \\
  &&A_2 = - \frac{\beta(1-\beta)(1+\beta)(1-2\beta)}{\kappa (-\eta)^2} \\
  &&A_3 = -  \frac{2\beta(1-\beta)(1+\beta)(1-2\beta)(1-11\beta)}{\kappa (-\eta)^3} \\
  &&A_4 =  \frac{3\beta(1-\beta)(1+\beta)^2 (1-2\beta)(1+3\beta)}{\kappa (-\eta)^4}
\end{eqnarray}
\begin{eqnarray}
  &&B_1 = -\frac{36\beta(1-\beta)(1+\beta)(1-2\beta)(1-3\beta)}{\sqrt{6} 
a_0 ~\kappa (-\eta)^{3+\beta}} \sqrt{\frac{1-\beta}{1-2\beta}} \\
  &&B_2 = \frac{4\beta(1-\beta)(1+\beta)(1-2\beta)(1-11\beta)}{\kappa 
(-\eta)^{3}}\\
  &&B_3 = \frac{12\beta(1-\beta)(1-2\beta)(1-3\beta)}{\kappa (-\eta)^2} \\
  &&B_4 = -\frac{48\beta(1-2\beta)^2 (1-\beta)}{\sqrt{6}~a_0~\kappa 
(-\eta)^{2+\beta} } \sqrt{\frac{1-\beta}{1-2\beta}}\\
  &&D_1 = \frac{9 \sqrt{(1-\beta)(1-2\beta)} (1+\beta)(1-\beta)
(3-5\beta + 5 \beta^2 -83 \beta^3)}{\sqrt{6}~a_0~\kappa (-\eta)^{4+\beta}}\\
  &&D_2 = \frac{36\beta\sqrt{(1-\beta)(1-2\beta)}(1-\beta)(1+\beta)
(7-17\beta)}{\sqrt{6}~a_0~\kappa (-\eta)^{3+\beta}}\\
  &&D_3 = - \frac{-12~a_0~\beta\sqrt{(1-\beta)(1-2\beta)}(1-2\beta)
(1-3\beta)}{\sqrt{6} \kappa (-\eta)^{2-\beta}}\\
  &&D_4 = \frac{72\beta\sqrt{(1-\beta)(1-2\beta)} 
(1-\beta)(1-2\beta)}{\sqrt{6}~a_0~\kappa (-\eta)^{2+\beta}}\\
  &&D_5 = \frac{12\beta(1-\beta)(1-2\beta)(1-3\beta)}{\kappa (-\eta)^2}\\
  &&D_6 = -\frac{24 \beta \sqrt{(1-2\beta)(1-\beta)}(1-2\beta)(1-\beta)
 }{\sqrt{6}~a_0~\kappa (-\eta)^{2+\beta}}\\
  &&E_1 = \frac{9\sqrt{(1-\beta)(1-2\beta)}(1-\beta)(1+\beta)^2(3-19\beta 
- 3\beta^2 -77\beta^3)}{\sqrt{6}~a_0~\kappa (-\eta)^{5+\beta}}\\
  &&E_2 = \frac{2~a_0~\sqrt{(1-\beta)(1-2\beta)}
(1+\beta)(1-2\beta)}{\sqrt{6}~\kappa (-\eta)^{1-\beta}}\\
  &&E_3 = -\frac{2~a_0~\sqrt{(1-\beta)(1-2\beta)}
(1+\beta)(1-2\beta)}{\sqrt{6}~\kappa (-\eta)^{1-\beta}}\\
  &&E_4 = \frac{72\beta\sqrt{(1-\beta)(1-2\beta)}(1-\beta)^2 
(1+\beta)(3+11\beta)}{\sqrt{6}~a_0~\kappa (-\eta)^{4+\beta}}\\
  &&E_5 = \frac{18\beta\sqrt{(1-\beta)(1-2\beta)}(1-\beta) 
(1+\beta)(7-17\beta)}{\sqrt{6}~a_0~\kappa (-\eta)^{3+\beta}}\\
  &&E_6 = -\frac{12~a_0~\beta\sqrt{(1-\beta)(1-2\beta)}(1-2\beta) 
(1+\beta)(1-3\beta)}{\sqrt{6}~\kappa (-\eta)^{3+\beta}}\\
  &&E_7 = \frac{4\beta(1-\beta)(1-2\beta)(1+\beta) 
(1-11\beta)}{\kappa (-\eta)^{3}}\\
  &&E_8 = - \frac{18\beta\sqrt{(1-\beta)(1-2\beta)}(1-\beta) 
(1+\beta)(1-3\beta)}{\sqrt{6}~a_0~\kappa (-\eta)^{3+\beta}}\\
\end{eqnarray}

After taking partial derivative,
\begin{equation}
\begin{split}
  & \int d^4x \Big[ \Big(A_1 \varphi_2^{\prime 2} +
  A_2 \partial^{i}{\varphi_2} \partial_{i}{\varphi_2} + A_5
  \varphi_2^2 \Big) + \varphi_2 ~ \Big(G_1 \varphi_1^2 + E_2 (\nabla^2
  B_1)^2 + E_3 \partial^{i j}{B_1} \partial_{i j}{B_1} + \\
  &
  G_2 \varphi_1
  \varphi_1^\prime + G_3 {\varphi_1^\prime}^2 + G_4 \varphi_1
  \varphi_1^{\prime \prime} + G_5 \varphi_1^\prime \varphi_1^{\prime
    \prime} + G_6 \varphi \nabla^2 \varphi_1 + G_7 \varphi^\prime
  \nabla^2 \varphi_1 +\\
  &
  G_8 \partial^{i}{\varphi_1}\partial_{i}{\varphi_1} +
  G_9 \partial^{i}{\varphi_1^\prime}\partial_{i}{\varphi_1} +
  G_{10} \partial^{i}{\varphi_1}\partial_{i}{B_1} +
  G_{11}\partial^{i}{\varphi_1^\prime}\partial_{i}{B_1} + G_{12} Q +
  G_{13} Q^\prime \Big) \Big]
  \end{split}
\end{equation}
where,
\begin{eqnarray}
  &&G_1 = E_1 - B_2 F_1 - D_1^\prime \\
  &&G_2 = E_4 - (B_2 F_2 + B_3 F_1) - (D_2^\prime + 2 D_1)\\
  &&G_3 = E_5 - B_3 F_2 - (D_2 + D_4^\prime)\\
  &&G_4 = - D_2 \\
  &&G_5 = -2D_4\\
  &&G_6 = -B_1\\
  &&G_7 = -B_4\\
  &&G_8 = E_8 - B_1 - (D_6^\prime - D_5 F_3)\\
  &&G_9 = - B_4 - 2 D_6\\
  &&G_{10} = E_7 - B_2 - (D_5^\prime - 2 \mathcal{H} D_5)\\
  &&G_{11} = -B_3 - D_5\\
  &&G_{12} = E_6 - D_3^\prime\\
  &&G_{13} = - D_3
\end{eqnarray}
so the equation of motion of scalar field for power law K-Inflation is,
\begin{equation}
\begin{split}
  &2 A_1 \varphi_2^{\prime \prime} + 2 A_1^\prime \varphi_2^{\prime} +
  A_2 \nabla^2{\varphi_2} + A_5 \varphi_2^2 = G_1 \varphi_1^2 + E_2
  (\nabla^2 B_1)^2 +
  E_3 \partial^{i j}{B_1} \partial_{i j}{B_1} + \\
  &
  G_2
  \varphi_1 \varphi_1^\prime +
  G_3 {\varphi_1^\prime}^2
  + G_4
  \varphi_1 \varphi_1^{\prime \prime} + G_5 \varphi_1^\prime
  \varphi_1^{\prime \prime} + G_6 \varphi \nabla^2 \varphi_1 + G_7
  \varphi^\prime \nabla^2 \varphi_1 +  \\
  &
  G_8 \partial^{i}{\varphi_1}\partial_{i}{\varphi_1} +
  G_9 \partial^{i}{\varphi_1^\prime}\partial_{i}{\varphi_1} +
  G_{10} \partial^{i}{\varphi_1}\partial_{i}{B_1} +
  G_{11}\partial^{i}{\varphi_1^\prime}\partial_{i}{B_1} + G_{12} Q +
  G_{13} Q^\prime
  \end{split}
\end{equation}

$a_0$ dependency in the action as well as in the equation of motion
can be resolved by rescaling $\varphi_1 \rightarrow a_0 \varphi_1,
\varphi_2 \rightarrow a_0 \varphi_2$.

\end{document}